\documentclass[nofootinbib,preprint,preprintnumbers,a4paper,10pt]{revtex4}
\usepackage{multirow}
\usepackage{textcomp}
\usepackage{amsmath,graphicx,color,epsfig}

\usepackage{footnote}
\usepackage{ulem}
\usepackage{booktabs}
\usepackage{array}
\usepackage{amssymb}
\usepackage{subfigure}
\usepackage{longtable}
\usepackage{verbatim}
\usepackage{amsfonts}
\usepackage{hyperref}
\usepackage{cancel}

\begin{document}
\title{CP asymmetries in $D\to K^0_{S,L}P$ and $D\to K^0_{S,L}V$ decays}
\author{Ying-Xin Lai$^{1}$ }
\author{Di Wang$^{1}$}\email{wangdi@hunnu.edu.cn}	
\address{$^1$Department of Physics, Hunan Normal University, Changsha 410081, China}

\begin{abstract}
$D$ meson decays into neutral kaons involve both Cabibbo-favored and doubly Cabibbo-suppressed amplitudes as well as final-state kaon mixing, providing abundant sources of CP violation.
In this work, we analyze CP asymmetries in the $D\to K^0_{S,L}P$ and $D\to K^0_{S,L}V$ decays, where $P$ and $V$ denote pseudoscalar and vector mesons respectively.
The formulas of the time-dependent and time-integrated CP asymmetries in these modes are derived, in which the $D^0-\overline D^0$ mixing effects and the $K^0_L$ modes are considered for the first time.
The hadronic parameters that determine CP asymmetries are extracted by the global fit of branching fractions within the topological diagram approach.
A significant result is that the tension between theoretical predictions and experimental data for the $K_S^0-K_L^0$ asymmetries in $D^0\to K_{S,L}^0\omega$ and $D^0\to K_{S,L}^0\phi$ modes is mitigated.
The CP-violating effects arising from the interference between Cabibbo-favored and doubly Cabibbo-suppressed amplitudes with neutral kaon mixing could reach to $\mathcal{O}(10^{-3})$ order in the $D^+\to K^0_S\pi^+$, $D^+_s\to K^0_SK^+$, $D^0\to K^0_S\rho^0$, and $D^0\to K^0_S\phi$ modes.
The difference between the CP asymmetries in the $D^+\to K^0_S\pi^+$ and $D^+_s\to K^0_SK^+$ modes will be available on LHCb and Belle II in the near future.

\end{abstract}
\maketitle
\section{Introduction}

CP violation (CPV) is a significant aspect of particle physics, as it is a crucial element in interpreting the matter-antimatter asymmetry in the Universe \cite{Sakharov:1967dj} and serves as a window for new physics.
It can be accommodated within the Standard Model (SM) by the Kobayashi-Maskawa (KM) mechanism \cite{Cabibbo:1963yz,Kobayashi:1973fv}.
CP violation in the charm sector is suppressed by the Glashow-Iliopoulos-Maiani (GIM) mechanism \cite{Glashow:1970gm}.
The LHCb Collaboration reported the discovery of the CP violation in the charm sector in 2019 \cite{Aaij:2019kcg},
\begin{equation}\label{z1}
  \Delta A_{ CP} \equiv A_{ CP}(D^0\to K^+K^-)-A_{ CP}(D^0\to \pi^+\pi^-) =(-1.54\pm0.29)\times 10^{-3}.
\end{equation}
In theory, it is difficult to precisely predict the CP asymmetries in the singly Cabibbo-suppressed $D$ meson decays owing to large uncertainties in penguin diagrams.
Those QCD-inspired approaches do not work well in charm decays because of the not large enough charm quark mass.
The penguin topologies cannot be extracted from branching fractions, because the CKM factors are much smaller those for tree topologies.

CP violation can also occur in $D$ meson decays into neutral kaons \cite{Azimov:1998sz,Amorim:1998pi,Bianco:2003vb,Bigi,Xing:1995jg,Lipkin:1999qz,DAmbrosio:2001mpr,Grossman:2011zk}. In these modes, the Cabibbo-favored (CF) and the doubly Cabibbo-suppressed (DCS) transitions are present simultaneously.
The interference between CF and DCS amplitudes, as well as neutral kaon mixing effect, induce measurable CP asymmetries.
The time-integrated CP asymmetry in the $D^+\rightarrow K_S^0\pi^+$ decay has been measured by the Belle Collaboration \cite{Ko:2012pe},
\begin{equation}\label{ex1}
A_{ CP}(D^+\rightarrow K_S^0\pi^+) = (-3.63\pm0.94\pm0.67)\times 10^{-3},
\end{equation}
with a $3.2\sigma$ deviation from zero.
Theoretical analyses show that CP asymmetries in $D$ decays into neutral kaons are dominated by the CP asymmetry in $K^0-\overline K^0$ mixing, $A_{CP}^{\overline K^0}\simeq -2\mathcal{R}e(\epsilon)\approx -3.23\times 10^{-3}$.
The direct CP asymmetry $A_{CP}^{\rm dir}$, which is induced by the interference between CF and DCS amplitudes, is smaller than $1\times 10^{-4}$.
In Ref.~\cite{Yu:2017oky}, we pointed out a new CP-violating effect, $A_{CP}^{\rm int}$, which is from the interference between the CF and DCS amplitudes with the mixing of final-state mesons.

In this work, we analyze the CP asymmetries in the $D\to K^0_{S,L}P$ and $D\to K^0_{S,L}V$ decays, where $P$ and $V$ are pseudoscalar and vector mesons, respectively.
The formulas of the time-dependent and time-integrated CP asymmetries in these modes are derived, in which the $D^0-\overline D^0$ mixing effect and the $K^0_L$ modes are taken into account.
The hadronic parameters that govern CP asymmetries are extracted within the topological diagram approach in which the Cabibbo-favored and doubly Cabibbo-suppressed modes are included in the global fit.
We find that the CP-violating effect $A_{CP}^{\rm int}$ could reach to $\mathcal{O}(10^{-3})$ order in several decay modes.
The difference in CP violation between the $D^+\to K^0_S\pi^+$ and $D^+_s\to K^0_SK^+$ modes is expected to be available on LHCb and Belle II in the near future.

This paper is organized as follows.
In Sec. \ref{for}, we formulate the time-dependent and time-integrated CP asymmetries for $D$ meson decays into neutral kaons.
In Sec. \ref{SM}, we predict the CP asymmetries for $D\to K^0_{S,L}P$ and $D\to K^0_{S,L}V$ modes within the topological diagram approach.
Section \ref{CON} is a short summary.
The CP transformations for the decay amplitudes are discussed in Appendix \ref{cp}.
\section{ Formalism }\label{for}

\subsection{Time-dependent CP violation}

In this section, we formulate the time-dependent and time-integrated CP asymmetries for the decays under consideration.
Under the convention $\mathcal{CP}|K^0\rangle = -|\overline{K}^0\rangle$, the neutral kaons $K_S^0$ and $K_L^0$ are expressed as linear combinations of $K^0$ and $\overline K^0$ \cite{PDG}
\begin{equation}\label{eq:KSKL}
|K_{S}^0\rangle  =   p|K^0\rangle-q|\overline{K}^0\rangle,\qquad
|K_{L}^0\rangle  =   p|K^0\rangle+q|\overline{K}^0\rangle,
 \end{equation}
where
\begin{equation}
p =   \frac{1+\epsilon}{\sqrt{2(1+|\epsilon|^2)}}, \qquad
q =   \frac{1-\epsilon}{\sqrt{2(1+|\epsilon|^2)}},
 \end{equation}
and $\epsilon$ is a complex parameter characterizing the CP asymmetry in neutral kaon mixing, with $|\epsilon|=(2.228\pm0.011)\times 10^{-3}$ and $\phi_\epsilon = 43.52^{\circ}\pm0.05^{\circ}$ \cite{PDG}.
For convenience, we denote the mass, width and lifetime of $K_{S,L}^0$ by $m_{S,L}$, $\Gamma_{S,L}$ and $\tau_{S,L}$, and $\Gamma = (\Gamma_S+\Gamma_L)/2$, $\Delta\Gamma=\Gamma_S-\Gamma_L$, $\Delta m = m_L-m_S$.

The $D\to \overline K^0f$ and $D\to K^0f$ decays are Cabibbo-favored (CF) and doubly Cabibbo-suppressed (DCS) transitions, respectively.
The amplitudes of these two modes can be written as
\begin{equation}\label{eq:ampCFDCS}
\begin{split}
\mathcal{A}(D\to \overline K^{0}f)=\mathcal{T}_{CF} e^{i(\phi_{CF}+\delta_{CF})},\qquad \mathcal{A}(D\to  K^{0}f)=\mathcal{T}_{DCS}\, e^{i(\phi_{DCS}+\delta_{DCS})},
\end{split}
\end{equation}
where $\mathcal{T}_{CF,\,DCS}$ are the magnitudes of the decay amplitudes, $\phi_{CF,\,DCS}$ are the weak phases, and $\delta_{CF,\,DCS}$ are the strong phases.
The amplitudes of the $\overline{D}\rightarrow K^0\overline{f}$ and $\overline{D}\rightarrow \overline K^0\overline{f}$ decays, where $\overline D$ and $\overline f$ denote the antiparticles of the $D$ and $f$ mesons, are given by
\begin{equation}\label{eq:ampbar}
  \mathcal{A}(\overline{D}\to K^0\overline{f})  = - \mathcal{T}_{CF}e^{i(-\phi_{CF}+\delta_{CF})},\qquad
  \mathcal{A}(\overline{D}\to \overline{K}^0\overline{f}) = - \mathcal{T}_{DCS}e^{i(-\phi_{DCS}+\delta_{DCS})}.
\end{equation}
The detailed derivation of Eq.~\eqref{eq:ampbar} is presented in Appendix \ref{cp}.
To express the formulas for CP asymmetries clearly, we write the ratio of the DCS to CF amplitudes as
 \begin{equation}\label{zq2}
 \mathcal{A}(D\rightarrow K^0f)/\mathcal{A}(D\rightarrow \overline{K}^0f) = r_f\,e^{i(\phi+\delta_f)},
 \end{equation}
where $r_f = \mathcal{T}_{DCS}/\mathcal{T}_{CF}$, $\phi=\phi_{DCS}-\phi_{CF}$ and $\delta_f = \delta_{DCS} - \delta_{CF}$.
The parameters $r_f$ and $\delta_f$ depend on the individual processes, but $\phi$ is mode independent in the SM, $\phi=Arg\left[-{V_{cd}^{*}V_{us}/V_{cs}^{*}V_{ud}}\right]=(-6.2\pm 0.4)\times 10^{-4}$ \cite{PDG}.
If we express the amplitudes of the $D\to K^0f$ and $D\to \overline K^0 f$ decays as
\begin{align}
\mathcal{A}(D\to K^0 f) &=   V^*_{cd}V_{us}\mathcal{T}^\prime_{DCS}e^{i\delta_{DCS}},  \quad
\mathcal{A}(D\to \overline K^0 f)    = V^*_{cs}V_{ud}\mathcal{T}^\prime_{CF}e^{i\delta_{CF}},
\end{align}
where $\mathcal{T}_{DCS,CF}^\prime$ donate the magnitudes of amplitudes excluding the CKM matrix elements, we get
 \begin{equation}\label{zq1}
 \frac{\mathcal{A}(D\rightarrow K^0f)}{\mathcal{A}(D\rightarrow \overline{K}^0f)} = \frac{ V^*_{cd}V_{us}}{V^*_{cs}V_{ud}}\frac{\mathcal{T}^\prime_{DCS}}{\mathcal{T}^\prime_{CF}}e^{i\delta_f}=-\Big|\frac{ V^*_{cd}V_{us}}{V^*_{cs}V_{ud}}\Big|\frac{\mathcal{T}^\prime_{DCS}}{\mathcal{T}^\prime_{CF}}e^{i(\phi+\delta_f)}\simeq-\tan^2\theta_C\hat r_f e^{i(\phi+\delta_f)},
 \end{equation}
where $\theta_C$ is the Cabibbo angle and $\tan^2\theta_C\simeq |( V^*_{cd}V_{us})/(V^*_{cs}V_{ud})|\sim\lambda^2\sim 0.05$, $\hat r_f=\mathcal{T}^\prime_{DCS}/\mathcal{T}^\prime_{CF}\sim \mathcal{O}(1)$, and then $r_f\simeq - \tan^2\theta_C \hat r_f\sim \mathcal{O}(10^{-2})$.

\begin{figure}[t!]
\includegraphics[scale=0.25]{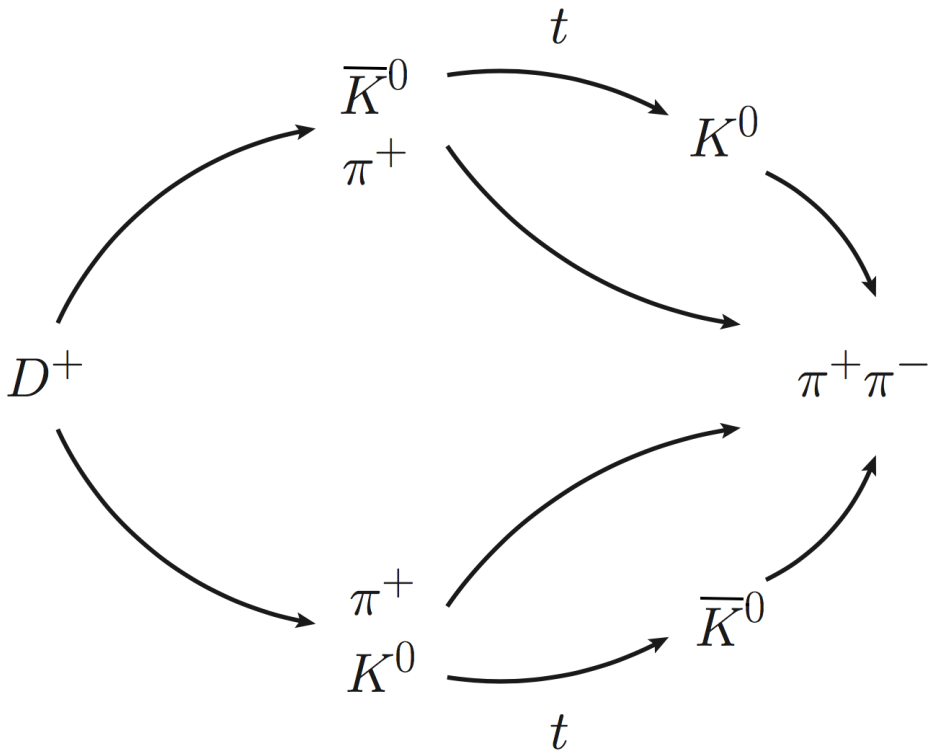}
\caption{Schematic description of the chain decay $D^+\to K(t)(\to \pi^+\pi^-)\pi^+$.
} \label{fig:amp}
\end{figure}

In experiments, a $K^0_S$ meson is reconstructed by $\pi^+\pi^-$ final state at a time close to its lifetime $\tau_S$.
Hence, both the $K^0_S$ and $K^0_L$ states serve as the intermediate states in the chain decays through the $K^0_S-K^0_L$ oscillation \cite{Grossman:2011zk}.
The chain decay, taking $D^+\to K^0_S\pi^+$ as an example, is depicted schematically in Fig.~\ref{fig:amp}.
The time-dependent CP asymmetry is defined by
\begin{equation}\label{m1}
A_{CP}(t) \equiv\frac{\Gamma_{\pi\pi}(t)-\overline
\Gamma_{\pi\pi}(t)}{\Gamma_{\pi\pi}(t)+\overline\Gamma_{\pi\pi}(t)},
\end{equation}
where
\begin{equation}
  \Gamma_{\pi\pi}(t)\equiv\Gamma(D\to K(t)(\to \pi^{+}\pi^{-})f),  \qquad
\overline\Gamma_{\pi\pi}(t)\equiv\Gamma(\overline D\to K(t)
(\to \pi^{+}\pi^{-})\overline f).
\end{equation}
The intermediate state $K(t)$ is a time-evolved neutral kaon, either $K^0(t)$ or $\overline K^0(t)$, and $t$ is the time difference between the charm decay and the neutral kaon decay in the kaon rest frame.
The amplitude $ \mathcal{A}(D\to K(t)(\to \pi^+\pi^-)f)$ can be written as
\begin{align}\nonumber
 & \mathcal{A}(D\to K(t)(\to \pi^+\pi^-)f)
   = \mathcal{A}(D\rightarrow K^0f)\big[g_+(t)\mathcal{A}(K^0\to \pi^+\pi^-)+\frac{q}{p}g_-(t)\mathcal{A}(\overline K^0\to \pi^+\pi^-)\big]\\\label{as1}
 &~~~~~~~~~~~~~~~~~~~~~~~~~~~~~~ +\mathcal{A}(D\rightarrow \overline K^0f)\big[g_+(t)\mathcal{A}(\overline K^0\to \pi^+\pi^-)+\frac{p}{q}g_-(t)\mathcal{A}(K^0\to \pi^+\pi^-)\big],
\end{align}
where $g_+$ and $g_-$ describe the flavor-preserving and flavor-changing time evolutions, respectively, and
\begin{equation}
  g_\pm(t)=\frac{1}{2}e^{-i (m_L -\frac{i}{2}\Gamma_L)t}\pm \frac{1}{2}e^{-i (m_S -\frac{i}{2}\Gamma_S)t}.
\end{equation}
The amplitude $ \mathcal{A}(\overline D\to K(t)(\to \pi^+\pi^-)\overline f)$ can be obtained by replacing $\mathcal{A}(D\rightarrow K^0f)$ and $\mathcal{A}(D\rightarrow \overline K^0f)$ with $\mathcal{A}(\overline D\rightarrow K^0\overline f)$ and $\mathcal{A}(\overline D\rightarrow \overline K^0\overline f)$ in Eq.~\eqref{as1}.

Neglecting the tiny direct CP asymmetry in $K^0\to \pi^+\pi^-$ decay, i.e., $\mathcal{A}(\overline K^0 \to \pi^+\pi^-)=-\mathcal{A}(K^0 \to \pi^+\pi^-)$, the time-dependent CP asymmetry is derived as
\begin{equation}\label{eq:KSAcp}
A_{CP}(t)\simeq\big[A_{CP}^{\overline K^0}(t)+A_{CP}^{\text{dir}}(t)+A_{CP}^{\text{int}}(t)\big]/D(t),
\end{equation}
with the denominator
\begin{equation}\label{de}
D(t)= e^{-\Gamma_St}(1-2\,r_f\cos\delta_f\cos\phi)+e^{-\Gamma_Lt}|\epsilon|^2.
\end{equation}
Here, we retain $|\epsilon|^2$ to avoid the singularity.
The first term $A_{CP}^{\overline K^0}(t)$ in numerator denotes the CP violation in neural kaon mixing \cite{Grossman:2011zk},
\begin{align}\label{q1}
A_{CP}^{\overline K^0}(t)
&=  2\mathcal{R}e(\epsilon)e^{-\Gamma_St}-2e^{-\Gamma t}
\big(\mathcal{R}e(\epsilon)\cos(\Delta mt)+\mathcal{I}m(\epsilon)\sin(\Delta mt)\big).
\end{align}
It is induced by the interference between CF decays with and without neutral kaon mixing.
It is independent of $r_f$, i.e., the DCS amplitude.
The second term $A_{CP}^{\text{dir}}(t)$ is the direct CP asymmetry induced by the interference between the CF and DCS amplitudes,
\begin{align}\label{q2}
A_{CP}^{\text{dir}}(t)=2e^{-\Gamma_St}\,r_f\sin\delta_f\sin\phi.
 \end{align}
The third term $A_{CP}^{\rm int}(t)$ arises from the interference between the CF and DCS amplitudes with neutral kaon mixing,
\begin{align}\label{q3}
A_{CP}^{\text{int}}(t)
&= -4\,r_f\sin\delta_f\cos\phi\big(\mathcal{I}m(\epsilon)e^{-\Gamma_St}-e^{-\Gamma t}
(\mathcal{I}m(\epsilon)\cos(\Delta mt)-\mathcal{R}e(\epsilon)\sin(\Delta mt))\big).
 \end{align}
It has been pointed out in \cite{Yu:2017oky} as a new CP-violating effect.
$A_{CP}^{\text{int}}(t)$ arises from the mother decay and the daughter mixing, which does not belong to any types of the traditional classification of CP violation given in PDG \cite{PDG}.
It is not the CP violation in neutral kaon mixing, as it vanishes in $r_f=0$ or $\delta_f=0$.
It is not the direct CP asymmetry in charm decays, as it does not vanish as $\phi\rightarrow 0$.
The mechanism responsible for $A^{\rm int}_{CP}(t)$ is different from the CP asymmetry induced by the interference between a decay without mixing $M^0\to f$ and a decay with mixing $M^0\to \overline M^0\to f$ in which both the oscillation and decay occur in the mother particle.
Its weak phase of $A^{\rm int}_{CP}(t)$ arises from (daughter) kaon mixing, $\epsilon$, while strong phase of $A^{\rm int}_{CP}(t)$ arises from (mother) charm decay, $\delta_f$.
To emphasize the specificity of $A^{\rm int}_{CP}(t)$, we compare the $A^{\rm int}_{CP}(t)$ with the three types of CPV given by PDG in Table~\ref{tab:cpv}.

\begin{table*}[t!]
\caption{Comparison of the CP-violating effect $A_{CP}^{\rm int}$ with the three CPV types given by PDG \cite{PDG}. }\label{tab:cpv}
\small
\begin{ruledtabular}
\begin{tabular}{|c|c|c|c|}
 & Source of width difference  & Weak phase & Strong phase  \\\hline
Direct CPV in decay &  $|\mathcal{A}(i\to f)|^2\neq |\mathcal{A}(\overline i\to \overline f)|^2$  &  Mother particle decay  &  Mother particle decay  \\\hline
Indirect CPV in mixing &  $|\mathcal{A}(P\to \overline P)|^2\neq |\mathcal{A}(\overline P\to P)|^2$  & Mother meson mixing  &   / \\\hline
CPV in interference between mother &  $|\mathcal{A}(P\to f)+\mathcal{A}(P\to\overline P\to f)|^2  $  &  Mother meson mixing  &  $\sin(\Delta mt)$, $\cos(\Delta mt)$  \\
 decay without and with mixing&  $\neq|\mathcal{A}(\overline P\to f)+\mathcal{A}(\overline P\to P\to f)|^2$  &   &    \\\hline
CPV in interference between &  $|\mathcal{A}(i\to P)+\mathcal{A}(i\to \overline P\to P)|^2$  &  Daughter meson mixing  & Mother particle decay \\
mother decay and daughter mixing & $\neq |\mathcal{A}(\overline i\to \overline P)+\mathcal{A}(\overline i\to P\to\overline P|^2$   &   &
\end{tabular}
\end{ruledtabular}
\end{table*}

The direct CP violation in charm decays is time-independent:
\begin{equation}
 \frac{A_{CP}^{\text{dir}}(t)}{D(t)} \simeq \frac{2\,r_f\sin\delta_f\sin\phi}{1-2\,r_f\cos\delta_f\cos\phi}.
\end{equation}
This is expected, since the direct CP asymmetry arises purely from charm decays and is independent of the time evolution of $K^0$ and $\overline K^0$.
Under the convention $\mathcal{CP}|K^0\rangle=-|\overline K^0\rangle$, the CP-even and CP-odd eigenstates of the neutral kaons are
\begin{equation}\label{zq3}
 |K_{+}^0\rangle  =  \big(|K^0\rangle-|\overline{K}^0\rangle\big)/\sqrt{2}, \qquad
|K_{-}^0\rangle  =   \big(|K^0\rangle+|\overline{K}^0\rangle\big)/\sqrt{2}.
\end{equation}
The direct CP asymmetry can be defined by
\begin{equation}\label{eq:AcpKplus}
A_{CP}(D\to K_+^0f) \equiv
     \frac{\Gamma(D\rightarrow K_+^0f)-\Gamma(\overline{D}\rightarrow K_+^0\overline{f})}{\Gamma(D\rightarrow K_+^0f)+\Gamma(\overline{D}\rightarrow K_+^0\overline{f})}.
\end{equation}
From Eqs. \eqref{q1} $\sim$ \eqref{q3}, it is found that $A_{CP}^{\overline K^0}(t)$ and $A_{CP}^{\text{int}}(t)$ vanish at $t=0$, and
\begin{equation}
  A_{CP}(t=0) = \frac{A_{CP}^{\text{dir}}(t=0)}{D(t=0)}= \frac{2\,r_f\sin\delta_f\sin\phi}{1-2\,r_f\cos\delta_f\cos\phi}.
\end{equation}
The direct CP asymmetry can thus be extracted by measuring the time-dependent CP asymmetry and extrapolating it to $t=0$.

\subsection{Time-integrated CP violation}\label{integral}

CP violation in the kaon system is governed by an efficiency function that reflects the specific experimental setup \cite{Grossman:2025uwz}.
If the efficiency function is neglected, the time-integrated CP asymmetry can be derived from the time-dependent asymmetry as
\begin{equation}\label{z21}
 A_{CP}(t_1,t_2)= \frac{\int^{t_2}_{t_1}dt\,\big[A_{CP}^{\overline K^0}(t)+A_{CP}^{\text{dir}}(t)+A_{CP}^{\text{int}}(t)\big]}
 {\int^{t_2}_{t_1}dt\,D(t)}.
\end{equation}
The time-integrated CP asymmetry reduces to
\begin{align}\label{int}
A_{CP}(t_1,t_2)&\simeq \frac{2r_f\sin\delta_f\sin\phi}{1-2r_f\cos\delta_f\cos\phi} + \frac{2\mathcal{R}e(\epsilon)-4\mathcal{I}m(\epsilon)\,r_f\cos\phi\sin\delta_f}{1-2r_f\cos\delta_f\cos\phi}\nonumber\\&~~~~~~~ \times \Bigg[1- \frac{\big[c(t_1)-c(t_2)\big]+\frac{\mathcal{I}m(\epsilon)+2\mathcal{R}e(\epsilon)\,r_f\cos\phi\sin\delta_f}{\mathcal{R}e(\epsilon)
-2\mathcal{I}m(\epsilon)\,r_f\cos\phi\sin\delta_f}
\big[s(t_1)-s(t_2)\big]}{\tau_S\Gamma \,(1+x^2)(e^{-t_1/\tau_S}-e^{-t_2/\tau_S})}\Bigg]
 \end{align}
with $x\equiv\Delta m/\Gamma$, $c(t)=e^{- \Gamma t}[\cos(\Delta m t)-x\, \sin(\Delta m t)]$, and $s(t)=e^{- \Gamma t}[x \cos(\Delta m t)+ \sin(\Delta m t)]$.
The first term, which is independent of $t_{1,2}$, corresponds to the direct CP asymmetry in charm decays.
In the rest of Eq. \eqref{int}, the terms proportional to $r_f$ represent the $A^{\rm int}_{CP}(t_1,t_2)$ term, and those independent of $r_f$ correspond to the CP violation in neutral kaon mixing.
In the limit $t_1\ll \tau_S\ll t_2 \ll \tau_L$, we have $e^{-\Gamma t_1}=e^{-\Gamma_S t_1}=1$ and $e^{-\Gamma t_2}=e^{-\Gamma_S t_2}=0$.
The time-integrated CP violation can be written as
\begin{align}\label{q4}
A_{CP}(t_1\ll \tau_S\ll t_2 \ll \tau_L)&\simeq \frac{2r_f\sin\delta_f\sin\phi}{1-2r_f\cos\delta_f\cos\phi}+ \frac{2(\mathcal{R}e(\epsilon)-2\mathcal{I}m(\epsilon)\,r_f\cos\phi\sin\delta_f)}{1-2r_f\cos\delta_f\cos\phi}\nonumber\\&~~~~~\times~~\Bigg[1-
\frac{2}{1+x^2}-\frac{\mathcal{I}m(\epsilon)+2\mathcal{R}e(\epsilon)\,r_f\cos\phi\sin\delta_f}
{\mathcal{R}e(\epsilon)-2\mathcal{I}m(\epsilon)\,r_f\cos\phi\sin\delta_f}\frac{2x}{1+x^2}\Bigg].
 \end{align}
Under the approximation of $\mathcal{R}e(\epsilon) / \mathcal{I}m(\epsilon) \simeq -y/x$ and $y\approx-1$ \cite{Grossman:2009mn}, we get
\begin{align}\label{x1}
 A_{CP}(t_1\ll \tau_S\ll t_2 \ll \tau_L)& \simeq
  \frac{-2\mathcal{R}e(\epsilon)+2r_f\sin\delta_f\sin\phi-4r_f\mathcal{I}m(\epsilon)\cos\phi\sin\delta_f}
 {1-2r_f\cos\phi\cos\delta_f}\nonumber\\& = \big[A_{CP}^{\overline K^0}+A_{CP}^{\text {dir}}+A_{CP}^{\text{int}}\big]/D.
\end{align}
The $K_{S}^{0}-K_{L}^{0}$ asymmetry is defined and expressed as
\begin{equation}\label{Rf}
   R(D\rightarrow K_{S,L}^0f)\equiv\frac{\Gamma(D\rightarrow K_S^0f) -\Gamma(D\rightarrow K_L^0f)}{\Gamma(D\rightarrow K_S^0f) + \Gamma(D\rightarrow K_L^0f)}\simeq -2\,r_{f}\cos\delta_f\simeq D-1.
 \end{equation}
In the SM, the total CP asymmetry is dominated by CP violation in neural kaon mixing, $A_{CP}^{\overline K^0}\simeq -2\mathcal{R}e(\epsilon)\approx -3.23\times 10^{-3}$.
The direct CP asymmetry $A_{CP}^{\rm dir}$ is of order of $10^{-5}$, since $r_f\sim \mathcal{O}(10^{-2})$ and $\phi\sim \mathcal{O}(10^{-4})$.
It is far below the precision of forthcoming experiments.
However, the CP-violating effect $A_{CP}^{\rm int}$ is of order of
$\mathcal{O}(10^{-4})$, which is significantly larger than the direct CP asymmetry.

Note that the interference between charm decay and $K^0-\overline K^0$ mixing, $A_{CP}^{\rm int}(t)$, is proportional to $r_f$ and $\sin\delta_f$.
This implies that the interference occurs only when both $D\to K^0f$ and $D\to \overline K^0f$ decays are present simultaneously.
In the SCS decay modes, such as $D^+\to K^0_SK^+$ and $D^+_s\to K^0_S\pi^+$, the $K^0_S$ meson originates only from $\overline K^0$ (for $D^+\to K^0_SK^+$) or $K^0$ (for $D^+_s\to K^0_S\pi^+$), but not both.
Due to the absence of two decay amplitudes with distinct origins, the interference between charm decays and $K^0-\overline K^0$ mixing vanishes.

In this work, we use the convention $\mathcal{CP}|K^0\rangle = -|\overline{K}^0\rangle$.
In some literature, such as Ref.~\cite{Xing:1995jg}, $\mathcal{CP}|K^0\rangle = +|\overline{K}^0\rangle$ is adopted.
Under this convention, the formulas for $A_{CP}^{\rm int}$ and $A_{CP}^{\rm dir}$ acquire an overall minus sign.
On the other hand, the ratio $r_f = T_{DCS}/T_{CF}\simeq -\tan^2\theta_C\hat r_f = -\tan^2\theta_C (\mathcal{T}^\prime_{DCS}/\mathcal{T}^\prime_{CF})$ has the opposite sign under the two conventions.
If we perform a phase transformation on $K^0$ and $\overline K^0$ as
\begin{equation}\label{pt}
 |K^0 \rangle \Rightarrow e^{i\alpha}|K^0 \rangle, \qquad |\overline{K}^0 \rangle \Rightarrow e^{i\beta}|\overline{K}^0 \rangle,
\end{equation}
 we have
\begin{equation}
  \mathcal{CP}|K^0 \rangle = |\overline{K}^0 \rangle  ~~\Rightarrow ~~\mathcal{CP}|K^0 \rangle = e^{i(\beta-\alpha)}|\overline{K}^0 \rangle.
\end{equation}
The conventions $\mathcal{CP}| K^0\rangle =-|\overline{K}^0\rangle$ and $\mathcal{CP}| K^0\rangle =+|\overline{K}^0\rangle$ are related to each other via Eq.~\eqref{pt} if $(\beta-\alpha)=\pi$.
The ratio $\mathcal{A}(D \to K^0f)/\mathcal{A}(D \to\overline{ K}^0f)$ transforms as
\begin{align}\label{z4}
 \frac{\langle K^0 f|\mathcal{H}_w|D \rangle}{\langle \overline{K}^0 f|\mathcal{H}_w|D \rangle}~~
 \Rightarrow~~  \frac{e^{-i\alpha}\langle  K^0 f|\mathcal{H}_w|D \rangle}{e^{-i\beta}\langle \overline{K}^0 f|\mathcal{H}_w|D \rangle}= e^{i(\beta-\alpha)}\frac{\langle K^0 f|\mathcal{H}_w|D \rangle}{\langle \overline{K}^0 f|\mathcal{H}_w|D \rangle}.
\end{align}
Thus, $r_f$ is not invariant under such a phase transformation.
The condition $(\beta-\alpha) = \pi$ gives $r_f$ the opposite sign between the two conventions.
As a result, the phase convention of $|K^0\rangle$ and $|\overline K^0\rangle$ does not affect the final formulas of CP asymmetries.

\subsection{$D^0-\overline D^0$ mixing effect}
\begin{figure}[t!]
\includegraphics[scale=0.40]{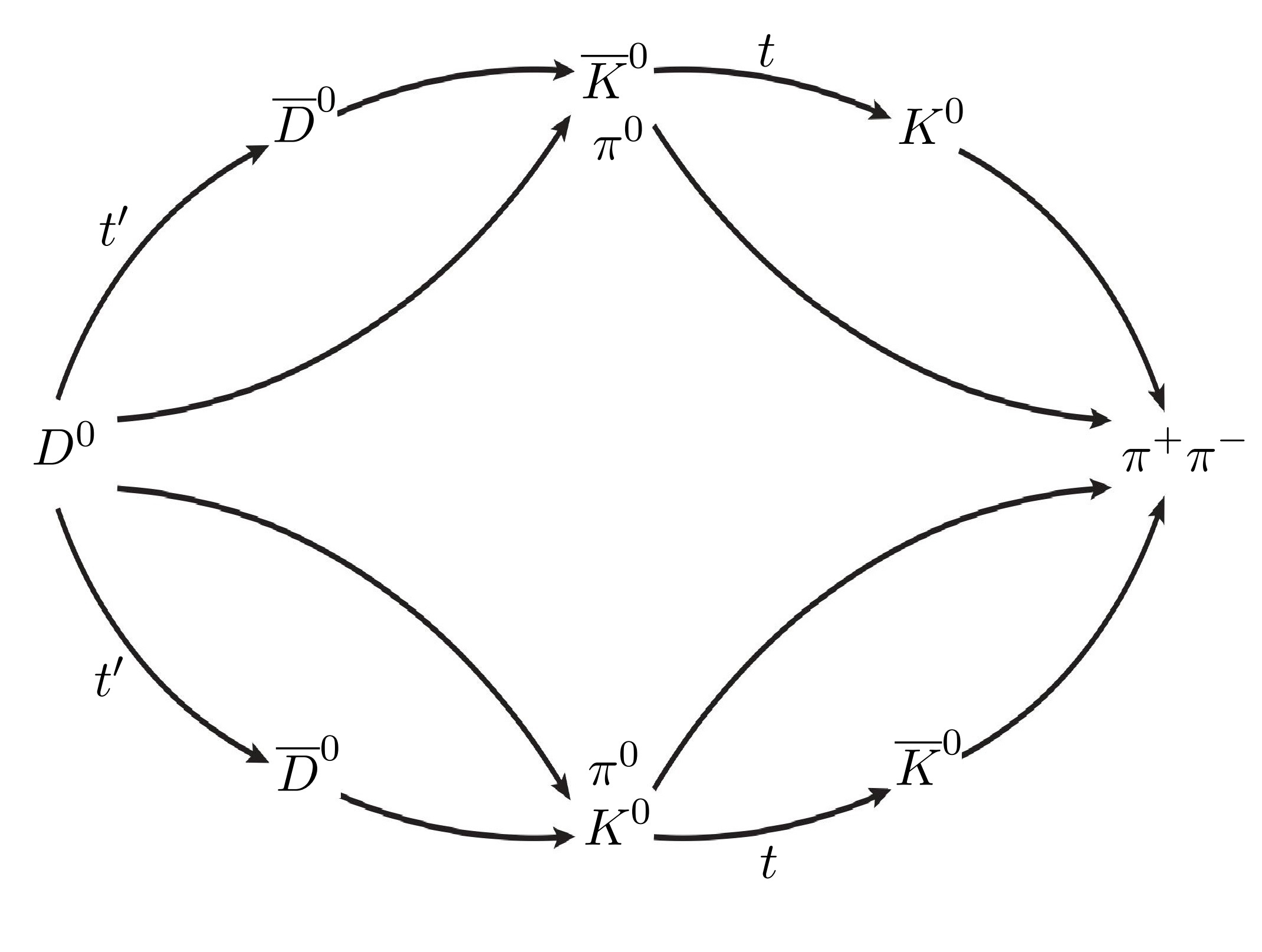}
\caption{Schematic description of the chain decay $D^0\to K^0_S(\to \pi^+\pi^-)\pi^0$.
} \label{fig:amp0}
\end{figure}

For $D^0$ meson decays into $K_{S}^0$ and a CP eigenstate $f_{CP}^0$, where $f_{CP}^0$ is $\pi^0$, $\eta$, $\eta'$, $\rho^0$, $\omega$ or $\phi$ in the $D\to PP$ and $D\to PV$ modes, the $D^0-\overline{D}^0$ mixing should be considered.
The chain decay $D^0\to K^0_S(\to \pi^+\pi^-)\pi^0$ is depicted schematically in Fig.~\ref{fig:amp0}.
To express the amplitudes of $D^0$ decays, we use the following standard notations:
\begin{align}\nonumber
   \mathcal{A}_{K_{S}^0}    & \equiv \mathcal{A}(D\to K(t)(\to \pi^+\pi^-)f), \quad \overline{\mathcal{A}}_{K_{S}^0}  \equiv \mathcal{A}(\overline D\to K(t)(\to \pi^+\pi^-)\overline f),
 \quad \lambda_{K_{S}^0}  \equiv \frac{q_D}{p_D}\frac{\overline{\mathcal{A}}_{K_{S}^0}}{\mathcal{A}_{K_{S}^0}}, \\
  \Gamma_{D^0}  & \equiv \frac{\Gamma_{D_1^0} + \Gamma_{D_2^0}}{2},  \quad
   x_D  \equiv  \frac{\Delta m_{D^0}}{\Gamma_{D^0}}  = \frac{m_{D_1^0} - m_{D_2^0}}{\Gamma_{D^0}}, \quad y_D \equiv \frac{\Delta \Gamma_{D^0}}{2 \Gamma_{D^0}} = \frac{\Gamma_{D_1^0} - \Gamma_{D_2^0}}{2\Gamma_{D^0}},
\end{align}
where $D_1^0$ and $D_2^0$ are the mass eigenstates of neutral $D$ mesons,
$|D^0_{1,2}\rangle = p_D|D^0\rangle \mp q_D|\overline{D}^0\rangle$ with $q_D/p_D = |q_D/p_D|e^{i\phi_D}$.
Here, the convention $\mathcal{CP}|D^0\rangle=-|\overline D^0\rangle$ is used.
Note that the final states of the $D^0\to K^0_{\pm}f^0_{CP}$ and $\overline D^0\to K^0_{\pm}f^0_{CP}$ decays are CP eigenstates.
From Eq.~\eqref{zq3}, the amplitudes of $\overline D^0\to K^0_+f^0_{CP}$ and $\overline D^0\to K^0_-f^0_{CP}$ decays are expressed as
\begin{align}\label{qq4}
 \mathcal{A}(\overline{D}^0\rightarrow K_+^0f^0_{CP})  &   = -\eta_{f^0_{CP}}\,\eta_{K_{+}^{0}}(-1)^{\mathcal{J}}\left[\mathcal{T}_{\rm DCS} e^{i(-\phi_{ DCS}+\delta_{ DCS})} - \mathcal{T}_{ CF}e^{i(-\phi_{CF}+\delta_{CF})}\right]/\sqrt2, \\\label{zq4}
\mathcal{A}(\overline{D}^0\rightarrow K_-^0f^0_{CP})  &   =-\eta_{f^0_{CP}}\,\eta_{K_{-}^{0}}(-1)^{\mathcal{J}}\left[ \mathcal{T}_{\rm DCS} e^{i(-\phi_{ DCS}+\delta_{ DCS})} + \mathcal{T}_{ CF} e^{i(-\phi_{ CF}+\delta_{ CF})}\right]/\sqrt2,
\end{align}
where $\eta_{f^0_{CP}}$ and $\eta_{K^0_{\pm}}$ denote the CP quantum numbers of $f^0_{CP}$ and $K^0_{\pm}$ mesons, and $\mathcal{J}$ is the orbital angular momentum of final state.
The overall minus sign in Eqs.~\eqref{qq4} and \eqref{zq4} is from $\mathcal{CP}|D^0\rangle=-|\overline D^0\rangle$.
Since $\eta_{K^0_\pm}=\pm1$,
we can express amplitudes of the $\overline D^0\to K^0f^0_{CP}$ and $\overline D^0\to \overline K^0f^0_{CP}$ as
\begin{align}\nonumber
    \mathcal{A}(\overline{D}^0\rightarrow K^0f^0_{CP}) & = \frac{1}{\sqrt{2}} \big[\mathcal{A}(\overline{D}^0\rightarrow K_+^0f^0_{CP})+ \mathcal{A}(\overline{D}^0\rightarrow K_-^0f^0_{CP})\big]\\\label{zq6}
      &~~ =\eta_{f^0_{CP}}(-1)^{\mathcal{J}}\mathcal{T}_{ CF} e^{i(-\phi_{ CF}+\delta_{CF})},\\\nonumber
    \mathcal{A}(\overline{D}^0\rightarrow \overline K^0f^0_{CP}) & = \frac{1}{\sqrt{2}} \big[-\mathcal{A}(\overline{D}^0\rightarrow K_+^0f^0_{CP})+ \mathcal{A}(\overline{D}^0\rightarrow K_-^0f^0_{CP})\big]\\\label{zq7}
      &~~ =\eta_{f^0_{CP}}(-1)^{\mathcal{J}}\mathcal{T}_{ CF} e^{i(-\phi_{ DCS}+\delta_{ DCS})}.
\end{align}
Note that $\eta_{f^0_{CP}}=-1$, $\mathcal{J}=0$ for $f^0_{CP}$ being a pseudoscalar meson and $\eta_{f^0_{CP}}=+1$, $\mathcal{J}=1$ for $f^0_{CP}$ being a vector meson.
One can find that Eqs.~\eqref{zq6} and \eqref{zq7} are the same as Eq.~\eqref{eq:ampbar}.

The $D^0$ meson mixing parameter $x_D = \Delta m_{D^0} /\Gamma_{D^0}$ is much smaller than that for $B^0_{d,s}$ mesons.
It is difficult to measure the time-dependent CP violation of charm mixing in experiments.
We therefore consider the time-integrated CP violation in charm mixing.
The time-integrated decay rates for the $D^0\to K_{S}^0f_{CP}^0$ and $\overline D^0\to K_{S}^0f_{CP}^0$ decays can be expressed as \cite{Grossman:2006jg}
\begin{align}\nonumber
  \Gamma(D^0\to K_{S}^0f_{CP}^0)  =& \int_0^\infty \Gamma(D^0(t^\prime)\to K_{S}^0f_{CP}^0)dt^\prime
   = |\mathcal{A}_{K_{S}^0}|^2 \Big[1+{1+|\lambda_{K_{S}^0}|^2\over2}\frac{y^2_D}{1-y^2_D} \\\label{eq:GammaD0KS}&- {1-|\lambda_{K_{S}^0}|^2\over2}\frac{x^2_D}{1+x^2_D}
       + \mathcal{R}e(\lambda_{K_{S}^0})\frac{y_D}{1-y^2_D} - \mathcal{I}m(\lambda_{K_{S}^0})\frac{x_D}{1+x^2_D}\Big],
\\\nonumber
  \Gamma(\overline{D}^0\to K_{S}^0f_{CP}^0)    = &\int_0^\infty \Gamma(\overline{D}^0(t^\prime)\to K_{S}^0f_{CP}^0)dt^\prime = |\overline{\mathcal{A}}_{K_{S}^0}|^2\Big[1+{1+|\lambda_{K_{S}^0}^{-1}|^2\over2}\frac{y^2_D}{1-y^2_D} \\\label{eq:GammaD0barKS} &- {1-|\lambda_{K_{S}^0}^{-1}|^2\over2}\frac{x^2_D}{1+x^2_D}
       + \mathcal{R}e(\lambda_{K_{S}^0}^{-1})\frac{y_D}{1-y^2_D} - \mathcal{I}m(\lambda_{K_{S}^0}^{-1})\frac{x_D}{1+x^2_D}\Big].
\end{align}
Here, $t^\prime$ is the time interval from $D^0$ production to its decay, which has been integrated over.
Note that the first terms in the above expressions are independent of $D^0-\overline{D}^0$ mixing, while the remaining terms are functions of the mixing parameters.

With the small values of $r_f$, $|\epsilon|$, $x_D$ and $y_D$, the total CP asymmetry in $D^0$ decays is derived as
\begin{equation}\label{eq:AcpD0KS}
A_{CP}(t)\simeq\big[A_{CP}^{\overline K^0}(t)+A_{CP}^{\text{dir}}(t)+A_{CP}^{\text{int}}(t)+ A_{CP}^{\overline D^0}(t) +  A_{CP}^{\rm dm }(t)+ A_{CP}^{ \overline D^0\rm,DCS  }(t) \big]/D_{D^0}(t),
\end{equation}
in which
\begin{align}\nonumber
D_{D^0}(t)=  & e^{-\Gamma_St}\Big(1-2\,r_f\cos\delta_f\cos\phi+\frac{1}{2}\Big(\Big|\frac{q_D}{p_D}\Big|+
\Big|\frac{p_D}{q_D}\Big|\Big)y_D\cos\phi_D\Big)+e^{-\Gamma_Lt}|\epsilon|^2\\&~~
\simeq e^{-\Gamma_St}\Big(1-2\,r_f\cos\delta_f\cos\phi+y_D\Big)+e^{-\Gamma_Lt}|\epsilon|^2,
\end{align}
and $t$ denotes the time interval from $D$ meson decay to neutral kaon decay.
The first three terms in numerator are the same as those in Eq.~\eqref{eq:KSAcp}.
The fourth term in numerator is the CP violation in $D^0-\overline D^0$ mixing,
\begin{equation}\label{56}
  A_{CP}^{\overline D^0}(t)=e^{-\Gamma_St}\Big(\frac{y_D}{2}\Big(\Big|\frac{q_D}{p_D}\Big|-\Big|\frac{p_D}{q_D}\Big|\Big)\cos\phi_D
  -\frac{x_D}{2}\Big(\Big|\frac{q_D}{p_D}\Big|+\Big|\frac{p_D}{q_D}\Big|\Big)\sin\phi_D\Big).
\end{equation}
After integrating over the flying time of neutral kaons, $A_{CP}^{\overline D^0}$ is derived as
\begin{equation}\label{D0}
  A_{CP}^{\overline D^0}=\frac{y_D}{2}\Big(\Big|\frac{q_D}{p_D}\Big|-\Big|\frac{p_D}{q_D}\Big|\Big)\cos\phi_D
  -\frac{x_D}{2}\Big(\Big|\frac{q_D}{p_D}\Big|+\Big|\frac{p_D}{q_D}\Big|\Big)\sin\phi_D = -a^{\rm m}_{CP}-a^{\rm i}_{CP} = - A^{\rm ind}_{CP},
\end{equation}
in which $a^{\text{m}}_{CP}$ is the indirect CP violation in $D^0-\overline D^0$ mixing and $a^{\text{i}}_{CP}$ is the interference between $D^0$ decay with and without $D^0-\overline D^0$ mixing \cite{Grossman:2006jg},
\begin{equation}\label{p1}
  a^{\text{m}}_{CP}=-\frac{y_D}{2}\Big(\Big|\frac{q}{p}\Big|-\Big|\frac{p}{q}\Big|\Big)\cos\phi_D,\qquad a^{\text{i}}_{CP}=\frac{x_D}{2}\Big(\Big|\frac{q}{p}\Big|+\Big|\frac{p}{q}\Big|\Big)\sin\phi_D.
\end{equation}
$A_{CP}^{\overline D^0}$ given by HFAG \cite{HFLAV} is
\begin{align}
A_{CP}^{\overline D^0}= -a^{\rm ind}_{CP} = (0.010\pm 0.012)\%,
\end{align}
which is much smaller than the CP violation in neutral kaon mixing.
The fifth term, $A_{CP}^{\rm dm}(t)$, is the interference between $D^0-\overline D^0$ mixing and $K^0-\overline K^0$ mixing,
\begin{align}\nonumber
A_{CP}^{\rm dm}(t)
&= \Big(y_D\Big(\Big|\frac{q_D}{p_D}\Big|-\Big|\frac{p_D}{q_D}\Big|\Big)\sin\phi_D
  +x_D\Big(\Big|\frac{q_D}{p_D}\Big|+\Big|\frac{p_D}{q_D}\Big|\Big)\cos\phi_D\Big)\mathcal{I}m(\epsilon)e^{-\Gamma_St} \\\nonumber
&~~~~ - \Big(y_D\Big(\Big|\frac{q_D}{p_D}\Big|-\Big|\frac{p_D}{q_D}\Big|\Big)\sin\phi_D
  +x_D\Big(\Big|\frac{q_D}{p_D}\Big|+\Big|\frac{p_D}{q_D}\Big|\Big)\cos\phi_D\Big)\mathcal{I}m(\epsilon)\cos(\Delta mt)e^{-\Gamma t}  \\
&~~~~~ + \Big(y_D\Big(\Big|\frac{q_D}{p_D}\Big|-\Big|\frac{p_D}{q_D}\Big|\Big)\sin\phi_D
  +x_D\Big(\Big|\frac{q_D}{p_D}\Big|+\Big|\frac{p_D}{q_D}\Big|\Big)\cos\phi_D\Big)\mathcal{R}e(\epsilon)\sin(\Delta mt)e^{-\Gamma t} .
\end{align}
It is a case of double-mixing induced $CP$ asymmetry pointed out in Refs.~\cite{Shen:2023nuw,Song:2025lmj}.
$A_{CP}^{\rm dm}(t)$ is doubly suppressed by the small parameters $x_D$, $y_D$ and $\epsilon$ and can be neglected.
The sixth term, $A_{CP}^{\overline D^0,\rm DCS}(t)$, is the interference between the DCS amplitudes and the CF amplitudes with the $D^0-\overline D^0$ mixing,
\begin{align}\nonumber
  A_{CP}^{\overline D^0,\rm DCS}(t)&=-e^{-\Gamma_St}r_f\bigg(\Big(\Big|\frac{q_D}{p_D}\Big|+\Big|\frac{p_D}{q_D}\Big|\Big)x_D\sin\phi
  \\  &~~~~~+\Big(y_D\Big(\Big|\frac{q_D}{p_D}\Big|-\Big|\frac{p_D}{q_D}\Big|\Big)\cos\phi_D
 -x_D\Big(\Big|\frac{q_D}{p_D}\Big|+\Big|\frac{p_D}{q_D}\Big|\Big)\sin\phi_D\Big)\bigg),
\end{align}
which is negligible in the SM.
The phase convention for $D^{0}$ and $\overline D^{0}$ does not affect the physical observables.
If the convention $\mathcal{CP}|D^{0}\rangle = + |\overline D^{0}\rangle$ is adopted, an additional minus sign would appear in Eqs.~\eqref{zq6},~\eqref{zq7}, as well as the $\mathcal{R}e(\lambda_{K_{S}^{0}})$ and $\mathcal{I}m(\lambda_{K_{S}^{0}})$ terms in Eqs.~\eqref{eq:GammaD0KS},~\eqref{eq:GammaD0barKS}.
Then Eq.~\eqref{eq:AcpD0KS} is the same under $\mathcal{CP}|D^{0}\rangle = + |\overline D^{0}\rangle$ and $\mathcal{CP}|D^{0}\rangle = -|\overline D^{0}\rangle$.

\subsection{CP violation in $K^0_L$ modes}\label{KL}

In Belle experiments, the pure $|K^0_L\rangle$ state can be established on $K^0_L$
and $\mu$ detection (KLM).
To formulate the CP violation in $K^0_L$ modes, a reasonable expression for $\langle K^0_L|$ state is required.
In the case of CP conservation, we have $\langle K^0_L|=(|K^0_L\rangle)^\dag$.
However, this equation is no longer valid in the presence of CP violation.
In general, a beam of oscillating and decaying neutral kaons can be described by a two-component wave function in its rest frame,
\begin{equation}
  |\psi(t)\rangle = \psi_1(t)|K^0\rangle + \psi_2(t)|\overline K^0\rangle,
\end{equation}
where $t$ is the proper time.
The wave function evolves according to the Schr\"{o}dinger-like equation:
\begin{equation}
i\frac{\partial}{\partial t}|\psi(t)\rangle = \mathcal{H}|\psi(t)\rangle.
\end{equation}
The $2\times 2$ matrix $\mathcal{H}$ is non-Hermitian.
Otherwise, the neutral kaons would only oscillate but not decay.
Without loss of generality, $\mathcal{H}$ can be written as
\begin{equation}
  \mathcal{H}=M-\frac{i}{2}\Gamma =
  \left(
   \begin{array}{cc}
     M_{11}-\frac{i}{2}\Gamma_{11}, & M_{12}-\frac{i}{2}\Gamma_{12} \\
       M_{21}-\frac{i}{2}\Gamma_{21}, & M_{22}-\frac{i}{2}\Gamma_{22} \\
        \end{array}
         \right),
\end{equation}
where $M=M^\dagger$, $\Gamma=\Gamma^\dagger$.
If CP is conserved, $\mathcal{H}$ commutes with its Hermitian conjugate.
The matrix $\mathcal{H}$ can be diagonalized by a unitary transformation.
However, in the presence of CP violation, $\mathcal{H}$ cannot be diagonalized by a unitary transformation, but rather by a general similarity transformation,
\begin{equation}
  X^{-1}\mathcal{H}X = \rm diag(\mu_H, \mu_L),
\end{equation}
where $X^{-1}\neq X^{\dagger}$.
Since $X$ is nonunitary, the left-eigenvectors of $\mathcal{H}$, $\langle K^0_S|$, and $\langle K^0_L|$-are not the Hermitian conjugates of their right eigenvectors, $|K^0_S\rangle$ and $| K^0_L\rangle$.
In this case, $\langle K^0_S|$ and $\langle K^0_L|$ are obtained from $X^{-1}$ rather than $X^{\dagger}$ \cite{Branco:1999fs}.
If CPT is conservative, the reciprocal basis can be defined as \cite{Branco:1999fs}
\begin{equation}
  \langle K^0_L|K^0_S\rangle = \langle K^0_S|K^0_L\rangle =0, \qquad   \langle K^0_L|K^0_L\rangle = \langle K^0_S|K^0_S\rangle =1.
\end{equation}
With these equations, the left eigenvectors of $\mathcal{H}$ can be expressed as
\begin{align}
 \langle K^0_S| & = \frac{1}{2p}\langle K^0| - \frac{1}{2q}\langle \overline K^0| = \sqrt{\frac{1+|\epsilon|^2}{2}}\Big(\frac{1}{1+\epsilon}\langle K^0| - \frac{1}{1-\epsilon}\langle \overline K^0|\Big),\\\label{x2}
  \langle K^0_L| & = \frac{1}{2p}\langle K^0| + \frac{1}{2q}\langle \overline K^0| = \sqrt{\frac{1+|\epsilon|^2}{2}}\Big(\frac{1}{1+\epsilon}\langle K^0| + \frac{1}{1-\epsilon}\langle \overline K^0|\Big).
\end{align}
According to Eq.~\eqref{x2}, the amplitudes for the $D\to K^0_Lf$ and $\overline D\to K^0_L\overline f$ decays read as
\begin{align}\label{KLamp1}
\mathcal{A}(D\to K^0_Lf) &=   \sqrt{\frac{1+|\epsilon|^2}{2}}\Big(\frac{1}{1+\epsilon}\mathcal{T}_{DCS}\, e^{i(\phi_{DCS}+\delta_{DCS})} + \frac{1}{1-\epsilon}\mathcal{T}_{CF} e^{i(\phi_{CF}+\delta_{CF})}\Big),\\\label{KLamp}
\mathcal{A}(\overline D\to K^0_L\overline f) &=   -\sqrt{\frac{1+|\epsilon|^2}{2}}\Big(\frac{1}{1+\epsilon}\mathcal{T}_{CF} e^{i(-\phi_{CF}+\delta_{CF})} + \frac{1}{1-\epsilon}\mathcal{T}_{DCS}\, e^{i(-\phi_{DCS}+\delta_{DCS})}\Big).
\end{align}
With Eqs.~\eqref{KLamp1} and \eqref{KLamp}, along with the convention in Eq.~\eqref{zq2}, the CP asymmetry in the $D\to K^0_{L}f$ modes is derived as
\begin{align}\nonumber
A_{CP}(D\to K^0_{L}f) & =
     \frac{
     |\mathcal{A}(D\rightarrow K_{L}^0f)|^2-|\mathcal{A}(\overline{D}\rightarrow K_{L}^0\overline{f})|^2
     }{
     |\mathcal{A}(D\rightarrow K_{L}^0f)|^2+|\mathcal{A}(\overline{D}\rightarrow K_{L}^0\overline{f})|^2
     }\\\nonumber
 &~~\simeq \frac{2\mathcal{R}e(\epsilon)-2\,r_f\sin\delta_f\sin\phi+4\,r_f\mathcal{I}m(\epsilon)\cos\phi\sin\delta_f}
 {1+2\,r_f\cos\phi\cos\delta_f}\\\label{acpkl}
 &~~~~=-\big[A_{CP}^{\overline K^0}+A_{CP}^{\text {dir}}+A_{CP}^{\text{int}}\big]/D^\prime.
\end{align}

It can be found that all terms in the numerator are opposite to those in the time-integrated CP asymmetry for $K_S^0$ modes under the limitation of $t_1\ll \tau_S\ll t_2 \ll \tau_L$.
The direct CP asymmetry is induced by the interference between CF and DCS amplitudes.
From Eqs.~\eqref{eq:AcpKplus} and \eqref{zq3}, we have
\begin{equation}\label{eq:AcpKcp}
\frac{\Gamma(D\rightarrow K^0_-f)-\Gamma(\overline{D}\rightarrow K^0_-\overline{f})}{\Gamma(D\rightarrow K^0_+f)-\Gamma(\overline{D}\rightarrow K^0_+\overline{f})}
=-1,
\end{equation}
where the minus sign between the $K_L^0$ and $K_S^0$ modes is evident.
Similarly, the difference in the denominators between the $K^0_L$ and $K^0_S$ modes can also be be understood since
\begin{align}
\frac{\Gamma(D\rightarrow K^0_-f)+\Gamma(\overline{D}\rightarrow K^0_-\overline{f})}{\Gamma(D\rightarrow K^0_+f)+\Gamma(\overline{D}\rightarrow K_+^0\overline{f})}& =\frac{1+2\,r_f\cos\delta_f\cos\phi}{1-2\,r_f\cos\delta_f\cos\phi}.
\end{align}
As for $A_{CP}^{\overline K^0}$ and $A_{CP}^{\text{int}}$ terms, they are significantly different between the $K^0_S$ and $K^0_L$ modes, since $A_{CP}^{\overline K^0}$ and $A_{CP}^{\text{int}}$ in chain decay $D\to K(t)(\to \pi^+\pi^-)f$ are associated with $K^0_S-K^0_L$ oscillation, to which both $K^0_S$ and $K^0_L$ contribute.
If a pure $K^0_S$ is considered, we can define a fake $CP$ violation in the $D\to K^0_{S}f$ mode, which is derived as
\begin{align}\nonumber
A^{\rm Fake}_{CP}(D\to K^0_{S}f) & =
     \frac{
     |\mathcal{A}(D\rightarrow K_{S}^0f)|^2-|\mathcal{A}(\overline{D}\rightarrow K_{S}^0\overline{f})|^2
     }{
     |\mathcal{A}(D\rightarrow K_{S}^0f)|^2+|\mathcal{A}(\overline{D}\rightarrow K_{S}^0\overline{f})|^2
     }\\\label{k}
 &~~\simeq \frac{2\mathcal{R}e(\epsilon)+2\,r_f\sin\delta_f\sin\phi-4\,r_f\mathcal{I}m(\epsilon)\cos\phi\sin\delta_f}
 {1-2\,r_f\cos\phi\cos\delta_f}.
\end{align}
This result is consistent with Eq.~\eqref{eq:KSAcp} if only the $e^{-\Gamma_St}$ terms are retained.
The opposite sign in Eqs.~\eqref{acpkl} and \eqref{x1} originates from the $e^{-\Gamma t}$ terms to which $K^0_L$ contribute and the approximations of $\mathcal{R}e(\epsilon)/\mathcal{I}m(\epsilon) \simeq -y/x$ and $y\approx-1$.

For $D^0$ decays into $K_L^0$, the time-integrated decay rates can be obtained by  replacing $\mathcal{A}_{K_S^0}$, $\overline{\mathcal{A}}_{K_S^0}$, $\lambda_{K_S^0}$ with $\mathcal{A}_{K_L^0}$, $\overline{\mathcal{A}}_{K_L^0}$, $\lambda_{K_L^0}$, respectively.
Then total CP asymmetry in the $D^0\to K_L^0f_{CP}^0$ reads as
\begin{align}\label{eq:AcpD0KL}
A_{CP}(D^0\to K_L^0f_{CP}^0)\simeq
-\big[A_{CP}^{\overline K^0}+A_{CP}^{\text {dir}}+A_{CP}^{\text{int}}+ A_{CP}^{\overline D^0}\big]/D^\prime_{D^0},
\end{align}
where $D^\prime_{D^0}\simeq 1+2\,r_f\cos\delta_f\cos\phi-y_D$.
One can find that the term induced by $D^0-\overline D^0$ mixing in Eq.~\eqref{eq:AcpD0KL} has the opposite sign compared to that in Eq.~\eqref{eq:AcpD0KS}.
If we neglect the CP violation in $K^0-\overline K^0$ mixing and note that $\phi_{DCS}\simeq\phi_{CF}\approx 0$ in the SM, we get
\begin{equation}
 \lambda_{K_L^0} \equiv \frac{q_D}{p_D}\frac{\overline{\mathcal{A}}_{K_L^0}}{\mathcal{A}_{K_L^0}} \simeq \frac{q_D}{p_D}\frac{\overline{\mathcal{A}}_{K_-^0}}{\mathcal{A}_{K_-^0}} = -\frac{q_D}{p_D}\frac{\overline{\mathcal{A}}_{K_+^0}}{\mathcal{A}_{K_+^0}} \simeq  - \frac{q_D}{p_D}\frac{\overline{\mathcal{A}}_{K_S^0}}{\mathcal{A}_{K_S^0}} \equiv -\lambda_{K_S^0}.
\end{equation}
Thereby, those terms proportional to $x_D$ and $y_D$ (but not $x_D^2$ and $y_D^2$) in Eqs.~\eqref{eq:GammaD0KS} and \eqref{eq:GammaD0barKS} have the opposite sign between the $K^0_S$ and $K^0_L$ modes, which explains the minus sign between Eqs.~\eqref{eq:AcpD0KL} and \eqref{eq:AcpD0KS}.

\section{Phenomenological analysis}\label{SM}

\subsection{Global fit for topological diagrams}

In this work, we perform a global fit for the Cabibbo-favored and doubly Cabibbo-suppressed $D\to PP$ and $D\to PV$ decays within the topological diagram approach.
For the CF and DCS decays, all topological diagrams are at tree level.
The $D\to PP$ decays are dominated by the color-allowed tree diagram $T$, the color-suppressed tree diagram $C$, the $W$-exchange diagram $E$, and the $W$-annihilation diagram $A$.
For $D\to PV$ decays, there exist two different sets of topological diagrams, since the spectator quark of the $D$ meson can enter a pseudoscalar or vector meson.
A subscript of $P$ or $V$ is attached to the flavor amplitudes and the associated strong phases to denote respectively whether the spectator quark in the $D$ meson ends up in the pseudoscalar or vector meson.
A detailed presentation about the topological diagrams in $D\to PP$ and $D\to PV$ decays can be found in literature such as Ref.~\cite{Cheng:2010ry}.
In this work, the SCS modes are not used to extract the hadronic parameters, because flavor $SU(3)$ breaking effects are known to be significant in the SCS modes.
For instance, the difference between penguin diagrams $P_d$ and $P_s$ (where subscript denotes the $d$ or $s$ in the quark loop), $\Delta P = P_d - P_s$, could enhance the $SU(3)$ symmetry breaking in the SCS modes.
The $\eta-\eta^\prime$ mixing angle is defined as
\begin{align}
  \left(
    \begin{array}{c}
      \eta \\
      \eta^\prime \\
    \end{array}
  \right) =
  \left(
   \begin{array}{cc}
   \cos\phi & -\sin\phi \\
   \sin\phi & \cos\phi \\
   \end{array}
   \right)\left(
            \begin{array}{c}
              \eta_q \\
             \eta_s  \\
            \end{array}
          \right).
\end{align}
In this work, we take $\phi=40.4^{\circ}$ since the newest measurements by BESIII gave $\phi=(40.0\pm2.0\pm0.6)^{\circ}$ \cite{BESIII:2023ajr} and $\phi=(40.2\pm2.1\pm0.7)^{\circ}$ \cite{BESIII:2023gbn}, and $\phi=40.4^{\circ}$ is often used in the literature.

\begin{table*}[t!]
\caption{Representations of topological amplitudes and branching fractions for the $D\to K^0_{S,L}P$ decays. Our results are given in the last column, compared to the experimental data~\cite{PDG}. }\label{tab:BrPP}
\begin{ruledtabular}
\small
\begin{tabular}{cccccc}
 Modes &   ~Representation~    &~
$\mathcal{B}_{\rm exp}$~& ~$\mathcal{B}_{\rm th}$~\\\hline
$D^0\to K_S^0\pi^0$ &  ${1\over2}V_{cd}^*V_{us}(C-E)-{1\over2}V_{cs}^*V_{ud}(C-E)$   & (1.240$\pm $0.022)\%    & (1.244$\pm$0.014)\% \\
$D^0\to K_L^0\pi^0$ &$ {1\over2}V_{cd}^*V_{us}(C-E)+ {1\over2}V_{cs}^*V_{ud}(C-E)$  &(0.976$\pm $0.032)\%    & (0.964$\pm$0.011)\% \\
{$D^0\to K_S^0\eta$ } & \begin{tabular}{c} $ V_{cd}^*V_{us}[{1\over2}(C+E)\cos\phi_\eta -{1\over\sqrt2}E\sin\phi_\eta ]$ \\
 $~~~~~~~~~- V_{cs}^*V_{ud}[{1\over2}(C+E)\cos\phi_\eta -{1\over\sqrt2}E\sin\phi_\eta ]$ \end{tabular} & {(0.508$\pm$0.013)\%}   &{(0.522$\pm$0.004)\%}\\
$D^0\to K_L^0\eta$  & \begin{tabular}{c} $ V_{cd}^*V_{us}[{1\over2}(C+E)\cos\phi_\eta -{1\over\sqrt2}E\sin\phi_\eta ]$ \\
 $~~~~~~~~~+ V_{cs}^*V_{ud}[{1\over2}(C+E)\cos\phi_\eta -{1\over\sqrt2}E\sin\phi_\eta ]$ \end{tabular} &  (0.434$\pm$0.016)\%   & (0.405$\pm$0.003)\%  \\
$D^0\to K_S^0\eta'$ & \begin{tabular}{c} $ V_{cd}^*V_{us}[{1\over2}(C+E)\sin\phi_\eta +{1\over\sqrt2}E\cos\phi_\eta ]$ \\
 $~~~~~~~~~- V_{cs}^*V_{ud}[{1\over2}(C+E)\sin\phi_\eta +{1\over\sqrt2}E\cos\phi_\eta ]$ \end{tabular} & (0.951$\pm $0.032)\%    &(0.984$\pm$0.023)\%  \\
$D^0\to K_L^0\eta'$ &\begin{tabular}{c} $ V_{cd}^*V_{us}[{1\over2}(C+E)\sin\phi_\eta +{1\over\sqrt2}E\cos\phi_\eta ]$ \\
 $~~~~~~~~~+ V_{cs}^*V_{ud}[{1\over2}(C+E)\sin\phi_\eta +{1\over\sqrt2}E\cos\phi_\eta ]$ \end{tabular} &  (0.812$\pm$0.035)\%   &  (0.763$\pm$0.018)\%\\
$D^+\to K_S^0\pi^+$ &$ {1\over\sqrt2}V_{cd}^*V_{us}(C+A)- {1\over\sqrt2}V_{cs}^*V_{ud}(T+C)$ & (1.561$\pm $0.031)\%  &(1.527$\pm$0.027)\%  \\
$D^+\to K_L^0\pi^+$ &$ {1\over\sqrt2}V_{cd}^*V_{us}(C+A)+ {1\over\sqrt2}V_{cs}^*V_{ud}(T+C)$ & (1.46$\pm $0.05)\%  &(1.56$\pm$0.03)\%  \\
$D_s^+\to K_S^0K^+$ & $ {1\over\sqrt2}V_{cd}^*V_{us}(T+C)- {1\over\sqrt2}V_{cs}^*V_{ud}(C+A)$& (1.500$\pm $0.014)\%    &(1.501$\pm$0.020)\%  \\
$D_s^+\to K_L^0K^+$ & $ {1\over\sqrt2}V_{cd}^*V_{us}(T+C)+ {1\over\sqrt2}V_{cs}^*V_{ud}(C+A)$& (1.49$\pm $0.06)\%    &(1.52$\pm$0.02)\%  \\
\end{tabular}
\end{ruledtabular}
\end{table*}
\begin{table*}[t!]
 \caption{Same as Table~\ref{tab:BrPP} for the $D\to K^0_{S,P}V$ decays. }\label{tab:BrPV}
\begin{ruledtabular}
\small\begin{tabular}{ccccccc}
 Modes &   ~Representation~    &~
$\mathcal{B}_{\rm exp}$~& ~$\mathcal{B}_{\rm th}$~\\\hline
  $D^0\to K_S^0\rho^0$ & $ {1\over2}V_{cd}^*V_{us}(C_V-E_P)- {1\over2}V_{cs}^*V_{ud}(C_V-E_V)$& $(0.64^{+0.06}_{-0.08})\%$  &$(0.57\pm0.04)\%$ \\
  $D^0\to K_L^0\rho^0$  &$ {1\over2}V_{cd}^*V_{us}(C_V-E_P)+ {1\over2}V_{cs}^*V_{ud}(C_V-E_V)$&  & $(0.46\pm0.04)\%$ \\
  $D^0\to K_S^0\omega$  &$ {1\over2}V_{cd}^*V_{us}(C_V+E_P)- {1\over2}V_{cs}^*V_{ud}(C_V+E_V)$&$(1.11\pm0.06)\%$ & $(1.16\pm0.06 )\%$ \\
  $D^0\to K_L^0\omega$  &$ {1\over2}V_{cd}^*V_{us}(C_V+E_P)+ {1\over2}V_{cs}^*V_{ud}(C_V+E_V)$& $(1.16\pm0.04)\%$ & $(1.12\pm0.05)\%$\\
  $D^0\to K_S^0\phi$ & $ {1\over\sqrt2}V_{cd}^*V_{us}E_V- {1\over\sqrt2}V_{cs}^*V_{ud}E_P$& $(0.425\pm 0.029)\%$  &$(0.395\pm0.018)\%$\\
  $D^0\to K_L^0\phi$ &$ {1\over\sqrt2}V_{cd}^*V_{us}E_V+ {1\over\sqrt2}V_{cs}^*V_{ud}E_P$& $(0.407\pm 0.023)\%$ & $(0.419\pm0.018)\%$ \\
  $D^+\to K_S^0\rho^+$  &$ {1\over\sqrt2}V_{cd}^*V_{us}(C_V+A_P)- {1\over\sqrt2}V_{cs}^*V_{ud}(T_P+C_V)$& $(6.14^{+0.60}_{-0.35})\%$  &$(6.54\pm0.43)\%$ \\
  $D^+\to K_L^0\rho^+$  &$ {1\over\sqrt2}V_{cd}^*V_{us}(C_V+A_P)+ {1\over\sqrt2}V_{cs}^*V_{ud}(T_P+C_V)$&  &$(7.39\pm0.45)\%$ \\
  $D_s^+\to K_S^0K^{*+}$  &$ {1\over\sqrt2}V_{cd}^*V_{us}(T_P+C_V)- {1\over\sqrt2}V_{cs}^*V_{ud}(C_V+A_P)$& $(0.80\pm0.07)\%$  & $(0.84\pm0.05)\%$\\
  $D_s^+\to K_L^0K^{*+}$  &$ {1\over\sqrt2}V_{cd}^*V_{us}(T_P+C_V)+ {1\over\sqrt2}V_{cs}^*V_{ud}(C_V+A_P)$&   & $(1.22\pm0.06)\%$\\
\end{tabular}
\end{ruledtabular}
\end{table*}
\begin{table*}[htp]
\caption{Experimental branching fractions we have used in the global fit of the $D\to PP$ modes compared with our predictions.
All data are taken from PDG~\cite{PDG}.}\label{BrPP}
\begin{ruledtabular}
\small\begin{tabular}{ccc|ccc}
Modes & $\mathcal{B}_{\text{exp}}$& $\mathcal{B}_{\text{th}}$ & Modes & $\mathcal{B}_{\text{exp}}$& $\mathcal{B}_{\text{th}}$\\
\hline
$D^0\to \pi^+K^-$ &(3.945$\pm $0.030)$\%$&($3.939\pm0.036$)\% & $D_s^+\to \pi^+\eta$ &(1.686$\pm $0.027)$\%$ & ($1.673\pm 0.024$)\%\\
$D_s^+\to \pi^+\eta^{\prime}$ &(3.95$\pm $0.08)$\%$&($3.96\pm0.03$)\%&
 $D^0\to \pi^-K^+$ &(1.50$\pm $0.07)\textpertenthousand&(1.60$\pm 0.01$)\textpertenthousand\\
$D^+\to \pi^0K^+$ &(2.08$\pm $0.21)\textpertenthousand&(2.14$\pm 0.02$)\textpertenthousand& $D^+\to K^+\eta$ &(1.25$\pm$0.16)\textpertenthousand&(1.38$\pm 0.01$)\textpertenthousand\\
$D^+\to K^+\eta^{\prime}$ &(1.85$\pm$0.20)\textpertenthousand&(1.37$\pm 0.02$)\textpertenthousand&  &&\\
\end{tabular}
\end{ruledtabular}
\end{table*}
\begin{table*}[htp]
\caption{  Same as Table \ref{BrPP} for the $D\to PV$ decays.}\label{BrPV}
\begin{ruledtabular}
\small\begin{tabular}{ccc|ccc}
Modes & $\mathcal{B}_{\text{exp}}$& $\mathcal{B}_{\text{th}}$ & Modes & $\mathcal{B}_{\text{exp}}$& $\mathcal{B}_{\text{th}}$\\
\hline
$D^0\to \pi^+K^{*-}$ &($5.42\pm0.40$)\%& (6.01$\pm$0.32)\%& $D^0\to \pi^0\overline K^{*0}$ &($3.15\pm0.31$)\% &(3.05$\pm$0.10)\% \\
$D^0\to K^-\rho^+$ &($11.2\pm0.7$)\%& (11.4$\pm$0.8)\%& $D^0\to \eta\overline K^{*0}$ &($1.41\pm0.12$)\% &(1.52$\pm$0.11)\% \\
$D^+\to \pi^+\overline K^{*0}$ &($1.57\pm0.13$)\%&(1.61$\pm$0.15)\% & $D_s^+\to \pi^+\rho^0$&($0.0114\pm0.0016$)\% &(0.0118$\pm$0.0027)\% \\
$D_s^+\to \pi^+\omega$ &($0.193\pm0.030$)\%&(0.228$\pm$0.009)\% & $D_s^+\to \pi^+\phi$&($4.5\pm0.4$)\% &(3.6$\pm$0.2)\% \\
$D_s^+\to K^+\overline K^{*0}$ &($3.82\pm0.07$)\%& (3.83$\pm$0.09)\%& $D_s^+\to \eta\rho^+$&($8.9\pm0.8$)\% & (7.2$\pm$0.3)\%\\
$D_s^+\to \eta^{\prime}\rho^+$ &($5.8\pm1.5$)\%& ($5.5\pm0.2$)\%& $D^0\to \pi^-K^{*+}$ &($3.45^{+1.80}_{-1.02}$)\textpertenthousand& (3.42$\pm$0.25)\textpertenthousand \\
 $D^+\to \pi^+K^{*0}$ &($3.45\pm0.60$)\textpertenthousand&(2.59$\pm$0.06)\textpertenthousand&$D^+\to \pi^0K^{*+}$&($3.4\pm1.4$)\textpertenthousand &(5.4$\pm$0.2)\textpertenthousand\\
$D^+\to K^+\rho^0$ &($1.9\pm0.5$)\textpertenthousand&(1.4$\pm$0.1)\textpertenthousand & $D^+\to K^+\omega$& ($0.57^{+0.25}_{-0.21}$)\textpertenthousand&(0.84$\pm$0.05)\textpertenthousand
\\
$D^+\to K^+\phi$& ($0.090\pm 0.012$)\textpertenthousand&(0.071$\pm$0.003)\textpertenthousand& $D^+\to \eta K^{*+}$ & ($4.4^{+1.80}_{-1.50}$)\textpertenthousand&(2.3$\pm$0.1)\textpertenthousand\\
 $D_s^+\to K^+K^{*0}$& ($0.92\pm0.51$)\textpertenthousand&(0.19$\pm$0.02)\textpertenthousand\\
\end{tabular}
\end{ruledtabular}
\end{table*}
\begin{table*}[t!]
\caption{\label{tab:table1}Results for $K_S^0-K_L^0$ asymmetries compared with the factorization-assisted topological-amplitude (FAT) approach \cite{Wang:2017ksn}, the $SU(3)_F$ symmetric topological diagram (STD) approach \cite{Cheng:2024hdo}, and experimental data taken from \cite{BESIII:2022xhe,He:2007aj,BESIII:2019kfh}.}
\begin{ruledtabular}
\small\begin{tabular}{ccccccc}
  & $R(f)_{\text{FAT}}$ &$R(f)_{\text{STD}}$ & $R(f)_{\text{EXP}}$ & $R(f)_{\text{This Work}}$\\
  \hline
   $D^0\to K_{S,L}^0\pi^0$ & $0.113\pm0.001$ & $0.107\pm0.009$ & $0.108\pm0.035$ & $0.127\pm0.001$ \\
   $D^0\to K_{S,L}^0\eta$ & $0.113\pm0.001$ & $0.107\pm0.008$& $0.080\pm0.022$ &  $0.127\pm0.001$ \\
   $D^0\to K_{S,L}^0\eta^\prime$ & $0.113\pm0.001$ & $0.107\pm0.017$& $0.080\pm0.023$ & $0.127\pm0.001$ \\
  $D^+\to K_{S,L}^0\pi^+$ & $0.025\pm0.008$ & $-0.013\pm0.013$&$0.022\pm0.024$ & $-0.011\pm0.002$ \\
  $D_s^+\to K_{S,L}^0K^+$ & $0.012\pm0.006$& $-0.006\pm0.020$ & $-0.021\pm0.025$& $-0.005\pm0.001$ \\
   $D^0\to K_{S,L}^0\rho^0$ & $0.113\pm0.001$ & $0.083\pm0.050$ &  & $0.104\pm0.007$ \\
   $D^0\to K_{S,L}^0\omega$ & $0.113\pm0.001$ & $0.089\pm0.035$ & $-0.024\pm0.031$ &  $0.016\pm0.003$\\
   $D^0\to K_{S,L}^0\phi$ & $0.113\pm0.001$ & $0.041\pm0.029$ & $-0.001\pm0.047$ &  $-0.030\pm0.010$\\
   $D^+\to K_{S,L}^0\rho^{+}$ & $-0.037\pm0.011$ & $-0.058\pm0.055$ & & $-0.061\pm0.003$ \\
   $D^+_s\to K_{S,L}^0K^{*+}$ & $-0.070\pm0.032$ & $-0.159\pm0.028$ &  &$-0.181\pm0.009$
\end{tabular}
\end{ruledtabular}
\end{table*}

The topological amplitudes for the $D\to K^0_{S,L}P$ and $D\to K^0_{S,L}V$ decays are presented in Tables~\ref{tab:BrPP} and \ref{tab:BrPV}, respectively.
And the other CF and DCS modes that are included in global fit are listed in Tables~\ref{BrPP} and \ref{BrPV}.
In this work, we use 17 branching fractions to fit seven parameters in the $D\to PP$ modes.
To include partial $SU(3)_F$ breaking effects, we introduce a factor $f_K/f_\pi\sim 1.193$ \cite{PDG} in the DCS amplitudes, the reason being that the long-distance rescattering contributions in all topological diagrams can be derived by twisting of quark lines of the color-allowed tree diagram $T$ \cite{Wang:2021rhd}.
The $\pi^+$ and $K^+$ mesons are emitted via $T$ diagram in the CF and DCS decay modes, respectively.
The best-fit results (in units of $10^{-6}$ GeV) for the $D\to PP$ modes are
\begin{align}
  |T|&=3.147\pm 0.009, \qquad |C|= 2.541\pm 0.009,\qquad \phi_C= (-152.25\pm0.26)^\circ,\nonumber\\
 |E|&= 1.447\pm 0.024, \qquad \phi_E= (122.32\pm0.36)^\circ,\qquad |A|= 0.395\pm 0.009,\qquad \phi_A= (-10.30\pm1.97)^\circ,
\end{align}
with $\chi^2/.dnf=2.2$.
For the $D\to PV$ modes, we use 26 branching fractions to fit 15 parameters.
We do not introduce a $SU(3)_F$ symmetry-breaking factor for the $D\to PV$ modes, because different mesons are emitted via $T_V$ and $T_P$ diagrams.
The best-fit results (in units of $10^{-6}$ GeV) for the $D\to PV$ modes are
\begin{align}\label{fitPV}
  |T_V|&= 2.677\pm 0.066, \qquad |T_P|= 5.528\pm 0.105,\qquad \phi_{T_P}= (73.80\pm4.22)^\circ,\nonumber\\
 |C_V|&= 1.952\pm 0.041, \qquad \phi_{C_V}= (254.44\pm2.24)^\circ,\qquad |C_P|= 2.310\pm 0.025,\qquad \phi_{C_P}= (206.21\pm1.13)^\circ,\nonumber\\
|E_V|&= 1.967\pm 0.051, \qquad \phi_{E_V}= (-38.09\pm2.03)^\circ,\qquad |E_P|= 1.557\pm 0.033,\qquad \phi_{E_P}= (64.58\pm4.01)^\circ,\nonumber\\
|A_V|&= 0.542\pm 0.012, \qquad \phi_{A_V}= (-174.79\pm5.24)^\circ,\qquad |A_P|= 0.347\pm 0.014,\qquad \phi_{A_P}= (178.41\pm5.64)^\circ,
\end{align}
with $\chi^2/.dnf=3.1$.
The predicted branching fractions for the $D\to PP$ and $D\to PV$ modes are listed in Tables~\ref{tab:BrPP}, \ref{tab:BrPV}, \ref{BrPP}, and \ref{BrPV}, comparing with experimental data.
One can find the predictions are very consistent with the data for most decay modes, indicting excellent $SU(3)_F$ symmetry in the Cabibbo-favored and doubly Cabibbo-suppressed $D$ meson decays.

The predicted $K_S^0-K_L^0$ asymmetries are shown in Table~\ref{tab:table1}.
Our results are compared with those from the factorization-assisted topological-amplitude (FAT) approach \cite{Wang:2017ksn}, the $SU(3)_F$ symmetric topological diagram (STD) approach \cite{Cheng:2024hdo}, and experimental data.
It is found that our results mitigate the tension between theoretical predictions and experimental data for the $K_S^0-K_L^0$ asymmetries in the $D^0\to K_{S,L}^0\omega$ and $D^0\to K_{S,L}^0\phi$ decays.
In the FAT approach, the assumption $E_P=E_V$ is adopted.
It leads to the relation $R(D^0\to K^0_{S,L}\rho^0)=R(D^0\to K_{S,L}^0\omega)=R(D^0\to K_{S,L}^0\phi)$.
In Ref.~\cite{Cheng:2024hdo}, the relative strong phase between $E_P$ and $E_V$ is not large.
However, the fit result shown in \eqref{fitPV} suggests that the relative strong phase between $E_P$ and $E_V$ is nearly $\pi/2$.
The discrepancy arises that we take the formula
\begin{align}
  \Gamma(D\to PV) = \frac{p_c}{8\pi m^2_D}|M|^2
\end{align}
to express the partial decay width, while Ref.~\cite{Cheng:2024hdo} adopts
\begin{align}
  \Gamma(D\to PV) = \frac{p^3_c}{8\pi m^2_V}|\tilde{M}|^2,
\end{align}
where $M=\tilde{M}(\epsilon\cdot p_D)$.
Furthermore, we fit the decay amplitudes using the branching fractions of both CF and DCS modes, whereas only CF modes are used in Ref.~\cite{Cheng:2024hdo}.

\subsection{Numerical results for CP asymmetries}

With the global fit in the last subsection, we extract the parameters $r_f$ and $\delta_f$ in the $D\to K^0_{S,L}P$ and $D\to K^0_{S,L}V$ modes.
The branching fractions of $D$ decays are proportional to the cosine of the relative strong phases of decay amplitudes, $\cos\delta_f$.
Solutions with opposite signs of $\delta_f$ contribute equally to the branching fractions.
The CP asymmetries $A_{CP}^{\rm dir}$ and $A_{CP}^{\rm int}$ are proportional to $\sin\delta_f$, and therefore should yield two opposite results.
For convenience, we denote the values listed in Table.~\ref{tab:pa} by $S_1$, and those obtained by flipping the signs of all strong phases as $S_2$.

\begin{table*}[t!]
\caption{Numerical results of ratios between DCS and CF amplitudes and relative strong phases in the neutral kaons involving modes. }\label{tab:pa}
\begin{ruledtabular}
\small
\begin{tabular}{cccccc}
 & $D^+\to K_{S}^{0}\pi^+$    &$D^+_s\to K_{S}^{0}K^+$ &$D^+\to K_{S}^{0}\rho^+$&  $D^+_s\to K_{S}^{0}K^{*+}$ \\\hline
~$r_f$~&$0.0961\pm0.0010$&$0.0421\pm0.0004$&$0.0308\pm0.0014$ & $0.0924\pm0.0042$ \\
~$\delta_f$~ &$(86.77\pm0.62)^\circ$&$-(86.77\pm0.62)^\circ$&$-(8.40\pm7.34)^\circ$ & $(8.40\pm7.34)^\circ$\\\hline
 & $D^0\to K_{S}^{0}\pi^0(\eta^{(\prime)})$    &$D^0\to K_{S}^{0}\rho^0$ &$D^0\to K_{S}^{0}\omega$&  $D^0\to K_{S}^{0}\phi$ \\\hline
~$r_f$~&$0.0636\pm0.0001$&$0.0857\pm0.0036$&$0.0081\pm0.0015$ & $0.0674\pm0.0023$ \\
~$\delta_f$~ &$(180.00\pm0.00)^\circ$&$-(127.78\pm2.49)^\circ$&$(178.64\pm11.78)^\circ$ & $(77.33\pm4.49)^\circ$
\end{tabular}
\end{ruledtabular}
\end{table*}
\begin{figure}[t!]
\includegraphics[width=0.45\textwidth]{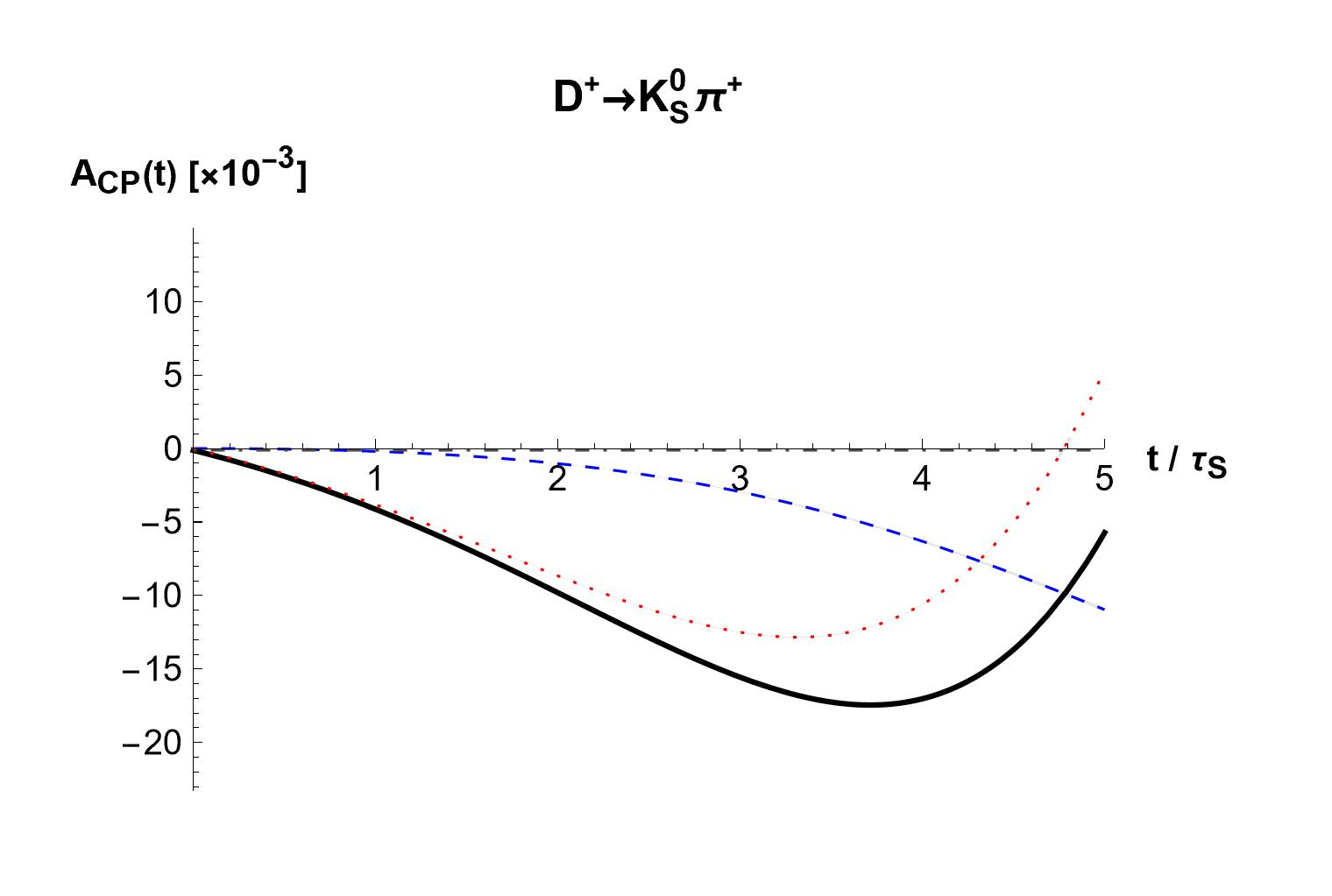}\qquad
\includegraphics[width=0.45\textwidth]{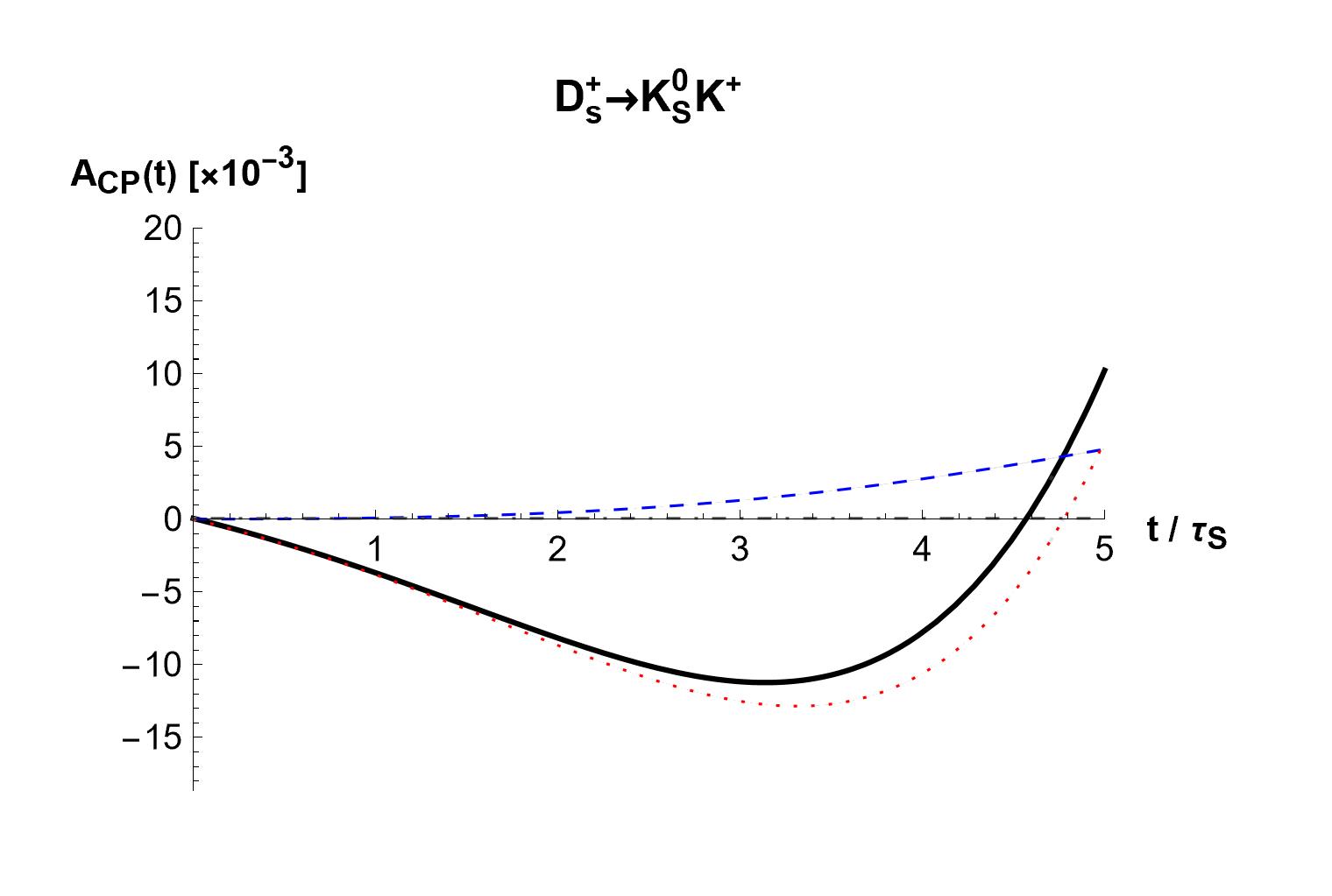}\\
\includegraphics[width=0.45\textwidth]{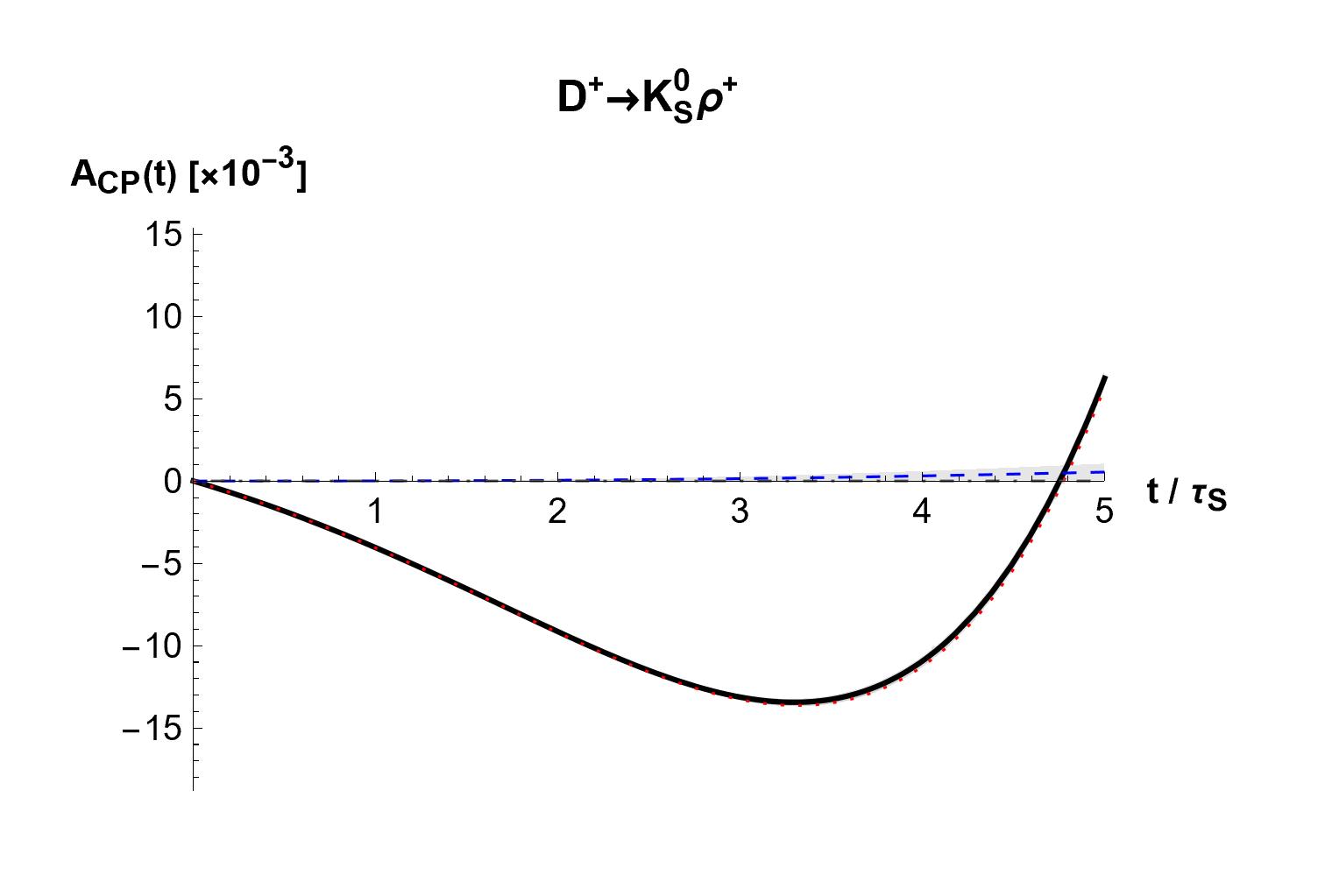}\qquad
\includegraphics[width=0.45\textwidth]{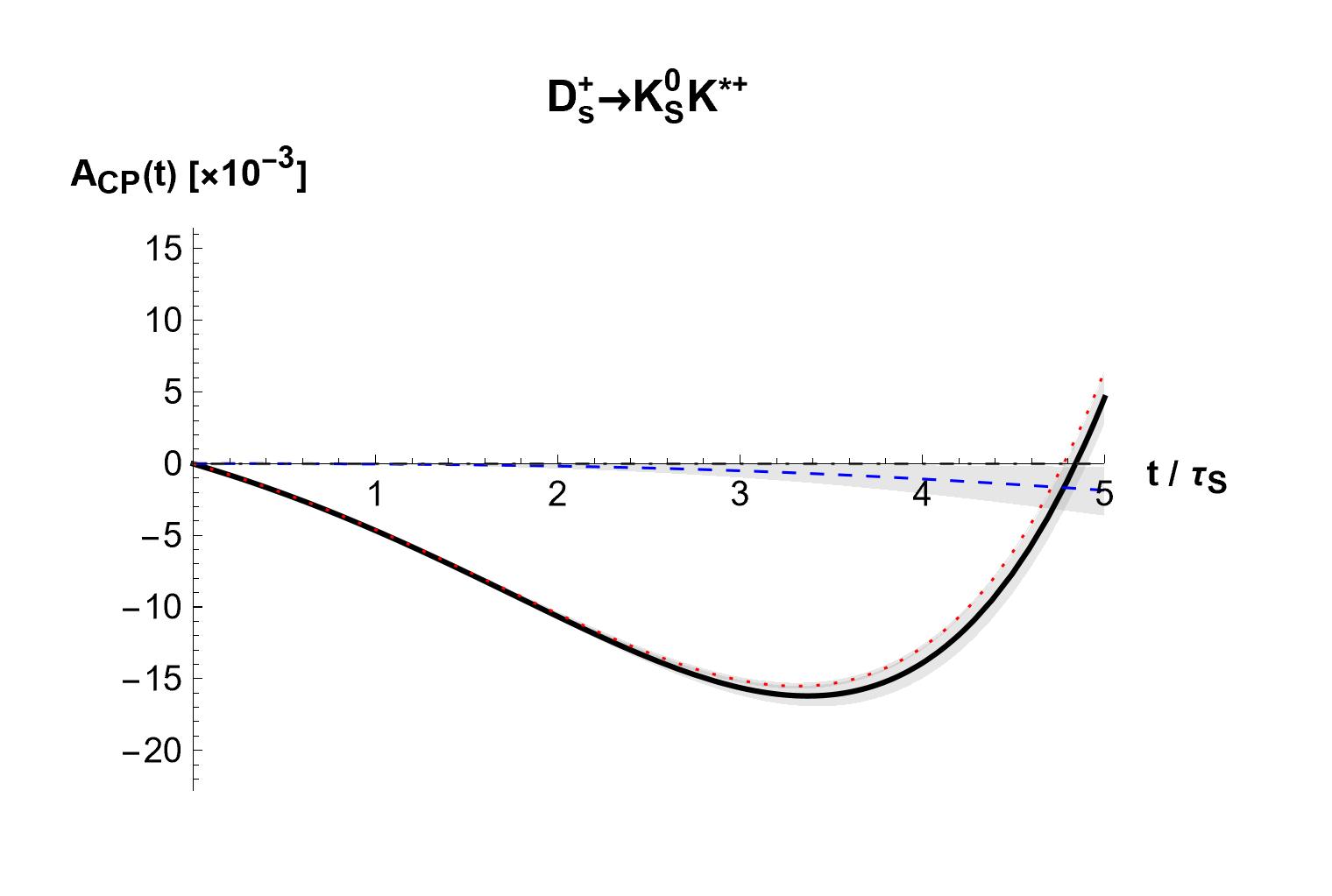}\\
\caption{Time-dependent CP asymmetries in the $D^+\to  K(t)(\to \pi^+\pi^-)\pi^+$, $D^+_s\to K(t)(\to \pi^+\pi^-)K^+$, $D^+\to K(t)(\to \pi^+\pi^-)\rho^+$, and $D^+_s\to K(t)(\to \pi^+\pi^-)K^{*+}$ decays as functions of $t/\tau_S$.
The red-dotted, gray dot-dashed, blue-dashed, and black-solid lines represent the $A_{CP}^{\overline K^0}(t)$, $A_{CP}^{\rm dir}(t)$, and $A_{CP}^{\rm int}(t)$ terms and total CP asymmetry $A_{CP}^{\rm tot}(t)$, respectively.
The gray bands represent the theoretical uncertainties.
The $A_{CP}^{\overline K^0}(t)$, $A_{CP}^{\rm dir}(t)$, and $A_{CP}^{\rm int}(t)$  have been normalized by denominator $D(t)$.
} \label{acpt1}
\end{figure}
\begin{figure}[t!]
\includegraphics[width=0.45\textwidth]{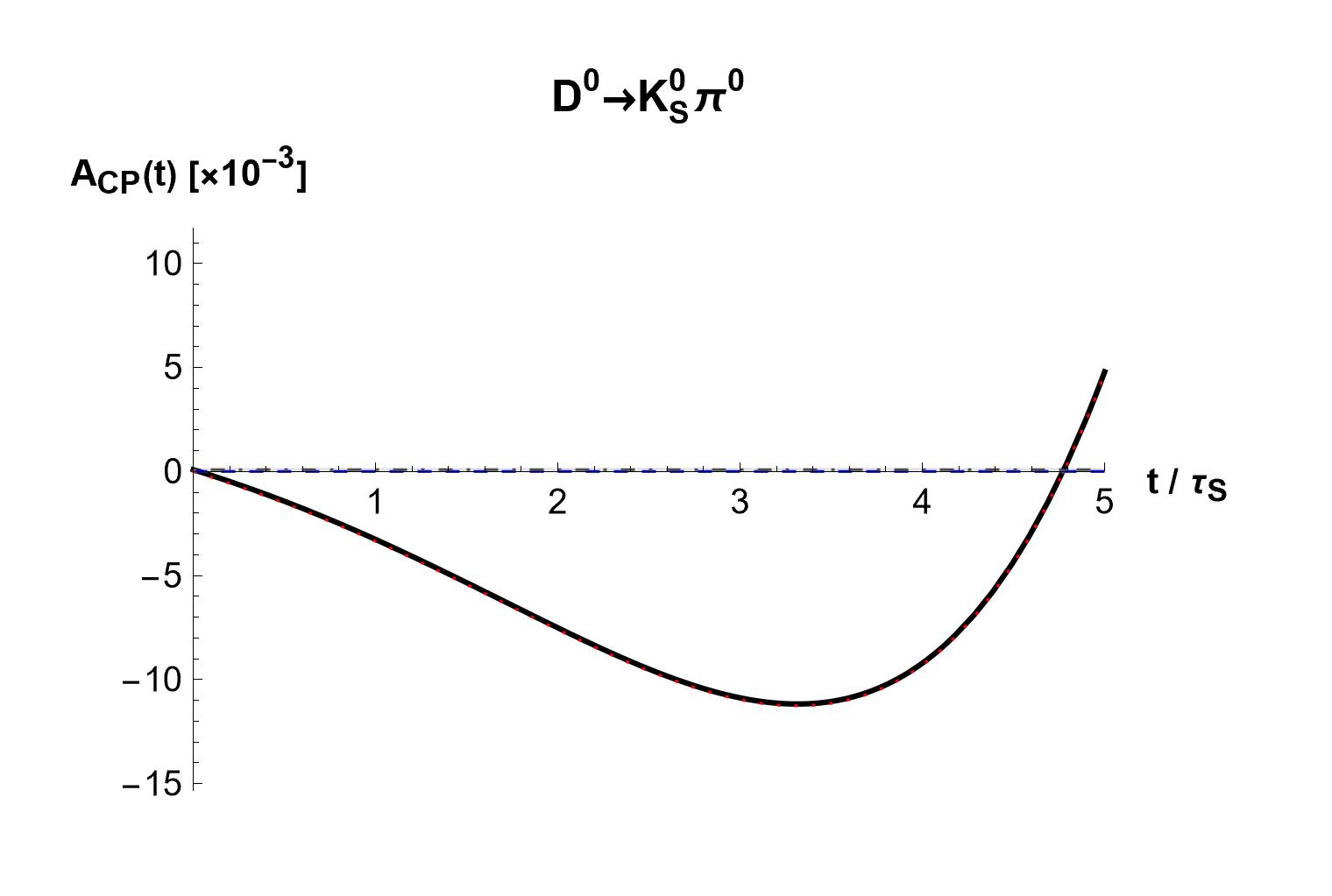}\qquad
\includegraphics[width=0.45\textwidth]{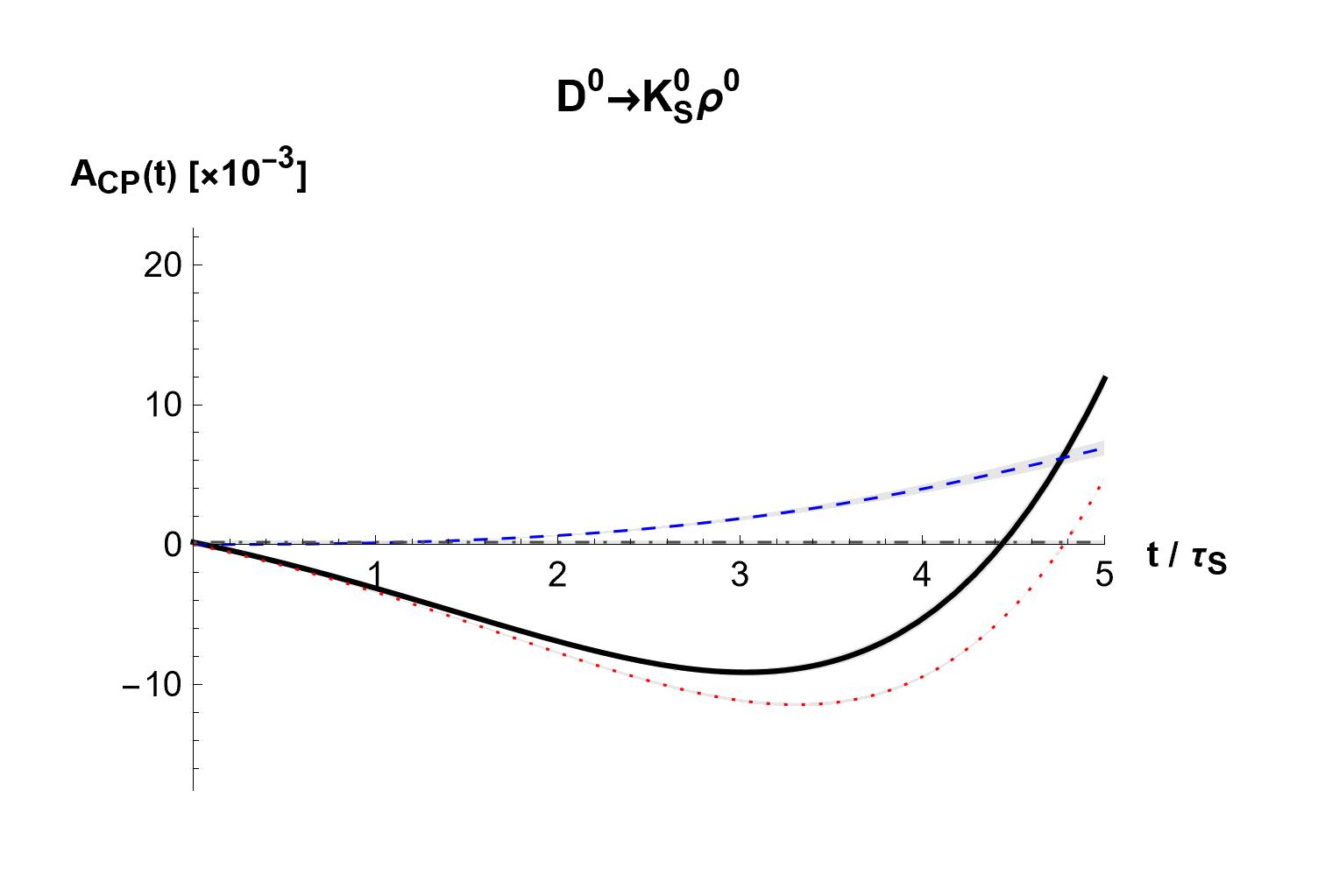}\\
\includegraphics[width=0.45\textwidth]{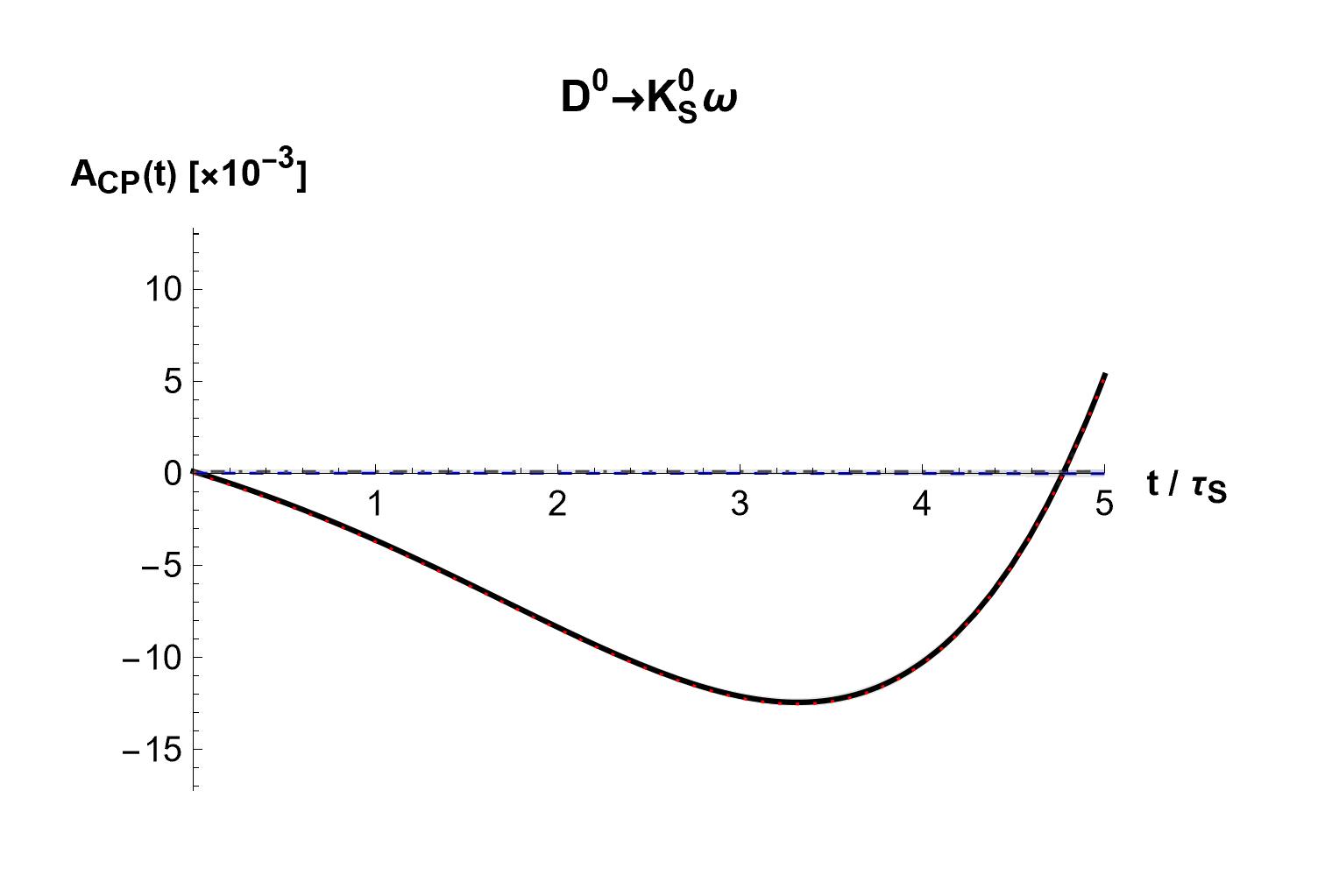}\qquad
\includegraphics[width=0.45\textwidth]{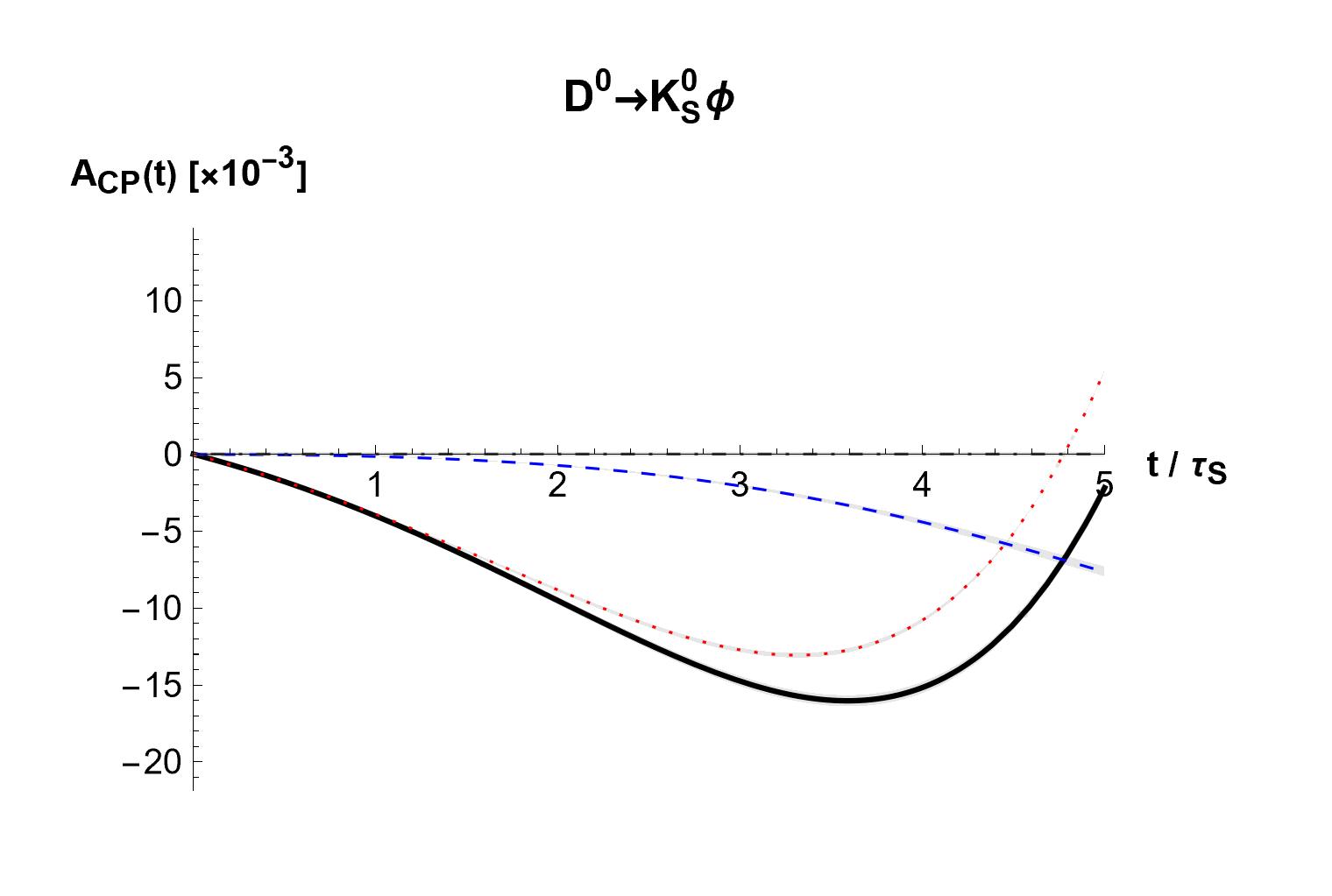}\\
\caption{Time-dependent CP asymmetries in the $D^0\to  K(t)(\to \pi^+\pi^-)\pi^0$, $D^0\to K(t)(\to \pi^+\pi^-)\rho^0$, $D^0\to K(t)(\to \pi^+\pi^-)\omega$, and $D^0\to K(t)(\to \pi^+\pi^-)\phi$ decays as functions of $t/\tau_S$, in which the gray dot-dashed line includes direct CP violation in charm decay $A_{CP}^{\rm dir}(t)$ and CP violation in $D^0-\overline D^0$ mixing $A_{CP}^{\overline D^0}(t)$.
} \label{acpt2}
\end{figure}

The time-dependent CP asymmetries in the $D\to  K(t)(\to \pi^+\pi^-)f$ decays as a function of $t/\tau_S$ are displayed in Figs.~\ref{acpt1} and \ref{acpt2} for charged and neutral $D$ meson decays, respectively.
Here, we only plot the results of $S_1$.
For $S_2$, the direct CP violation $A_{CP}^{\rm dir}(t)$ and the CP-violating effect $A_{CP}^{\rm int}(t)$ can be obtained by reflecting the results of $S_1$ with respect to time axis.
The gray bands in these figures represent the  associated uncertainties.
From Figs.~\ref{acpt1} and \ref{acpt2}, it is found that the total CP asymmetries are dominated by the $A_{CP}^{\overline K^0}(t)$ term.
The CP-violating effect $A_{CP}^{\rm int}(t)$ could reach to be of order of $10^{-3}$ in the $D^+\to K_S^0\pi^+$, $D^+_s\to K_S^0K^+$, $D^0\to K_S^0\rho^0$, and $D^0\to K_S^0\phi$ modes.

According to Eq.~\eqref{Rf}, the $K^0_S-K^0_L$ asymmetry reaches zero if $\delta_f\to \pm\pi/2$ or $r_f\to 0$.
From the fitting results, the small values of $R(D^+\to K^0_{S,L}\pi^+)$, $R(D^+_s\to K^0_{S,L}K^+)$, and $R(D^0\to K^0_{S,L}\phi)$ are attributed to the former case, while the small $R(D^0\to K^0_{S,L}\omega)$ is attributed to the latter.
According to Eq.~\eqref{q3}, the CP-violating term $A_{CP}^{\rm int}(t)$ is proportional to $\sin\delta_f$.
Thus, the CP-violating effect $A_{CP}^{\rm int}(t)$ is enhanced in the $D^+\to K^0_{S}\pi^+$, $D^+_s\to K^0_{S}K^+$, and $D^0\to K^0_{S}\phi$ modes.
The topological amplitudes contributing to the $D^0\to K^0\omega$ and $D^0\to K^0\rho^0$ decays are
 \begin{align}\label{x}
 \mathcal{A}(D^0\to K^0\omega)=\frac{1}{\sqrt{2}}V_{cd}^*V_{us}(C_V+E_P),\qquad
 \mathcal{A}(D^0\to K^0\rho^0)=\frac{1}{\sqrt{2}}V_{cd}^*V_{us}(C_V-E_P).
 \end{align}
The small value of $r_{\omega}$ is induced by the destructive interference between $C_V$ and $E_P$ in the $D^0\to K^0\omega$ mode.
According to Eq.~\eqref{x}, $C_V$ and $E_P$ interfere constructively in the $D^0\to K^0\rho^0$ mode and then $A_{CP}^{\rm int}(t)$ in $D^0\to K^0_S\omega$ is enhanced.

In the $SU(3)_F$ limit, the topological amplitudes contributing to the $D^+\to K^0\pi^+$,  $D^+\to\overline K^0\pi^+$, $D^+_s\to K^0K^+$, and $D^+_s\to\overline K^0K^+$ modes are
\begin{align}
\mathcal{A}(D^+\to K^0\pi^+)&=V^*_{cd}V_{us}(C+A),\qquad \mathcal{A}(D^+\to \overline K^0\pi^+)=V^*_{cs}V_{ud}(T+C),\nonumber\\
\mathcal{A}(D^+_s\to K^0K^+)&=V^*_{cd}V_{us}(T+C),\qquad \mathcal{A}(D^+_s\to \overline K^0K^+)=V^*_{cs}V_{ud}(C+A).
\end{align}
If we define $\mathcal{A}_1= C+A$ and $\mathcal{A}_2= T+C$, the ratios of the DCS to CF amplitudes in these two decay modes are
\begin{align}
 \frac{\mathcal{A}(D^+\to K^0\pi^+)}{\mathcal{A}(D^+\to \overline K^0\pi^+)}\simeq-\tan^2\theta_C \frac{\mathcal{A}_1}{\mathcal{A}_2},\qquad
 \frac{\mathcal{A}(D^+_s\to K^0K^+)}{\mathcal{A}(D^+_s\to \overline K^0K^+)}
 \simeq-\tan^2\theta_C \frac{\mathcal{A}_2}{\mathcal{A}_1}.
\end{align}
Then $r_{\pi^+}$ and $r_{K^+}$ can be written as
\begin{align}
r_{\pi^+}  = -\tan^2\theta_C
\times r,\qquad
r_{K^+}&= -\tan^2\theta_C\times \frac{1}{r},
\end{align}
where $r=\mathcal{A}_1/\mathcal{A}_2$.
The strong phases $\delta_{\pi^+}$ and $\delta_{K^+}$ are opposite under $U$-spin symmetry.
There are only three independent parameters-$|\mathcal{A}_1|$, $|\mathcal{A}_2|$, and the relative strong phase $\delta_{U}$ between $\mathcal{A}_1$ and $\mathcal{A}_2$-that determine the four branching fractions $\mathcal{B}r(D^+\to K^0_{S}\pi^+)$, $\mathcal{B}r(D^+\to K^0_{L}\pi^+)$, $\mathcal{B}r(D^+_s\to K^0_{S}K^+)$, and $\mathcal{B}r(D^+_s\to K^0_{L}K^+)$.
From Table \ref{tab:BrPP}, one can find that our fitting results are consistent with the experimental data, which indicates that the values of $|\mathcal{A}_1|$, $|\mathcal{A}_2|$, and $\delta_{U}$, and then the hadronic parameters $r_f$ and $\delta_f$ in the $D^+\to K^0_{S,L}\pi^+$ and $D^+_s\to K^0_{S,L}K^+$ modes, are well determined.

The topological amplitudes contributing to the $D^0\rightarrow K^0\pi^0$ and $D^0\rightarrow \overline{K}^0\pi^0$ decays are
\begin{equation}
 \mathcal{A}(D^0\rightarrow K^0\pi^0)=V^*_{cd}V_{us}(C+E),\qquad
 \mathcal{A}(D^0\rightarrow \overline{K}^0\pi^0)=V^*_{cs}V_{ud}(C+E).
\end{equation}
In the $SU(3)_F$ limit, we have $r_{\pi^0}=-\tan^2\theta_C$ and $\delta_{\pi^0}=0$.
Then, the $K^0_S-K^0_L$ asymmetry $R(D^0\to K^0_{S,L}\pi^0)$ is $0.107$ and the CP-violating effect $A_{CP}^{\rm int}(t)$ in $D^0\to K^0_{S}\pi^0$ is zero.
The experimental result for $R(D^0\to K^0_{S,L}\pi^0)$ is consistent with the $SU(3)_F$ prediction.
The CP-violating effect $A_{CP}^{\rm int}(t)$ in $D^0\to K^0_{S,L}\pi^0$ arises from $SU(3)_F$ breaking effects, which are predicted to be small.
Similar discussions can also be applied to the $D^0\to K^0_{S}\eta$ and $D^0\to K^0_{S}\eta^{\prime}$ modes.

\begin{table*}[t!]
\caption{Numerical results of the total CP asymmetries and the $A_{CP}^{\rm dir}$, $A_{CP}^{\rm int}$ terms for the $D\to K^0_Sf$ modes in the limit $t_1\ll\tau_S\ll t_2\ll\tau_L$ and the total CP asymmetries in the $D\to K^0_Lf$ modes (in units of $10^{-3}$). }\label{tab:acp}
\begin{ruledtabular}
\small
\begin{tabular}{cccccc}
 & $D^+\to K_{S}^{0}\pi^+$    &$D^+_s\to K_{S}^{0}K^+$ &$D^+\to K_{S}^{0}\rho^+$&  $D^+_s\to K_{S}^{0}K^{*+}$ \\\hline
~$A_{CP}^{\rm tot}(S_1)$~&$-3.98\pm0.01$&$-2.94\pm0.01$&$-3.41\pm0.04$ & $-4.08\pm0.10$ \\
~$A_{CP}^{\rm tot}(S_2)$~ &$-2.55\pm0.01$&$-3.56\pm0.01$&$-3.48\pm0.03$ & $-3.83\pm0.13$\\
\hline
~$A_{CP}^{\rm dir}(S_1)$~& $-0.120\pm0.001$&$0.052\pm0.001$&$0.006\pm0.005$ & $-0.020\pm0.018$\\
\hline
~$A_{CP}^{\rm int}(S_1)$~ &$-0.595\pm0.006$&$0.259\pm0.002$&$0.029\pm0.026$ & $-0.101\pm0.088$\\
\hline\hline
 & $D^+\to K_{L}^{0}\pi^+$    &$D^+_s\to K_{L}^{0}K^+$ &$D^+\to K_{L}^{0}\rho^+$&  $D^+_s\to K_{L}^{0}K^{*+}$ \\\hline
~$A_{CP}^{\rm tot}(S_1)$~&$3.90\pm0.01$&$2.91\pm0.01$&$3.01\pm0.03$ & $2.82\pm0.08$ \\
~$A_{CP}^{\rm tot}(S_2)$~ &$2.50\pm0.01$&$3.53\pm0.01$&$3.08\pm0.03$ & $2.65\pm0.07$\\\hline\hline
 & $D^0\to K_{S}^{0}\pi^0(\eta^{(\prime)})$    &$D^0\to K_{S}^{0}\rho^0$ &$D^0\to K_{S}^{0}\omega$&  $D^0\to K_{S}^{0}\phi$ \\\hline
~$A_{CP}^{\rm tot}(S_1)$~&$-2.76\pm0.12$&$-2.37\pm0.12$&$-3.06\pm0.12$ & $-3.70\pm0.13$ \\
~$A_{CP}^{\rm tot}(S_2)$~ &$-2.76\pm0.12$&$-3.27\pm0.12$&$-3.06\pm0.12$ & $-2.71\pm0.13$\\
\hline
~$A_{CP}^{\rm dir}(S_1)$~& $0$&$0.076\pm0.004$&$-0.000\pm0.002$ & $-0.083\pm0.003$\\
\hline
~$A_{CP}^{\rm int}(S_1)$~ &$0$&$0.374\pm0.020$&$-0.001\pm0.010$ & $-0.413\pm0.015$\\
\hline\hline
 & $D^0\to K_{L}^{0}\pi^0(\eta^{(\prime)})$    &$D^0\to K_{L}^{0}\rho^0$ &$D^0\to K_{L}^{0}\omega$&  $D^0\to K_{L}^{0}\phi$ \\\hline
~$A_{CP}^{\rm tot}(S_1)$~&$3.61\pm0.12$&$2.96\pm0.13$&$3.21\pm0.12$ & $3.53\pm0.13$ \\
~$A_{CP}^{\rm tot}(S_2)$~ &$3.61\pm0.12$&$4.09\pm0.13$&$3.20\pm0.12$ & $2.59\pm0.12$
\end{tabular}
\end{ruledtabular}
\end{table*}

The numerical results of the time-integrated CP asymmetries in $K^0_S$ modes in the limitation of $t_1\ll\tau_S\ll t_2\ll\tau_L$ and the CP asymmetries in $K^0_L$ modes are presented in Table~\ref{tab:acp}.
It can be seen that the total CP asymmetries for $S_1$ and $S_2$ are quite different.
The direct CP asymmetries in most modes are smaller than $1\times 10^{-4}$, while the CP-violating effect $A^{\rm int}_{CP}$ could be as large as several $10^{-4}$.
Our results for $A_{CP}(D^+\to K_S^0\pi^+)$ in two solutions are consistent with the experimental data given by Belle collaboration \cite{Ko:2012pe}, listed in Eq.~\eqref{ex1}.
Future measurements would help us to identify the sign of the strong phases.

\begin{figure}[t!]
\centering
\includegraphics[width=0.35\textwidth]{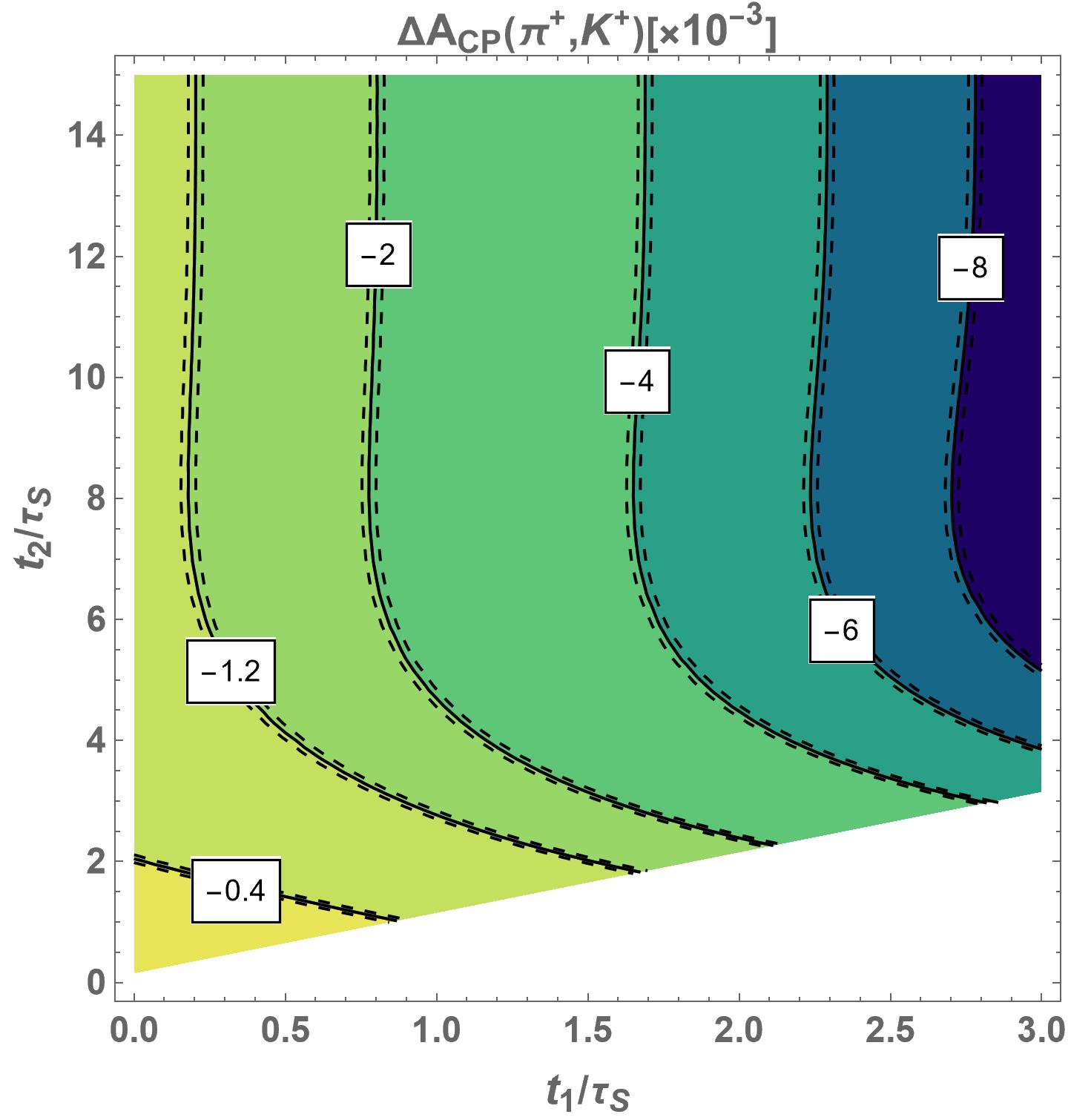}
\caption{The observable $\Delta A_{CP}(\pi^+,K^+)$ as a function of $t_1$ and $t_2$.}\label{fig:DeltaAcp}
\end{figure}

To verify the CP-violating effect induced by interference between CF and DCS amplitudes with neutral kaon mixing, we proposed an observable, the difference between the time-integrated CP asymmetries in the $D^+\to K^0_S\pi^+$ and $D^+_s\to K^0_SK^+$ modes in Ref.~\cite{Yu:2017oky},
\begin{equation}
  \Delta A_{CP}(\pi^+,K^+)\equiv A_{CP}(D^+\to K_S^0\pi^+)(t_1,t_2)-A_{CP}(D^+_s\to K_S^0K^+)(t_1,t_2).
\end{equation}
The CP violation in neutral kaon mixing is mostly eliminated in this observable.
Furthermore, the $A_{CP}^{\rm dir}$ and $A_{CP}^{\rm int}$ terms in the $D^+\to K^0_S\pi^+$ and $D^+_s\to K^0_SK^+$ modes have the opposite signs and are constructive in $\Delta A_{CP}(\pi^+,K^+)$.
By combining the three-body decays of $D^+\to K^-\pi^+\pi^+$ and $D^+_s\to K^+K^-\pi^+$, the dominant systematic asymmetries in measuring $\Delta A_{CP}(\pi^+,K^+)$, including the asymmetries in the selection of $\pi$ and $K$ mesons, the asymmetries in the production and discrimination different $D$ mesons, and the $K^0-\overline K^0$ asymmetries in materials, cancel each other out,
\begin{align}\nonumber
\Delta A_{CP}(\pi^+,K^+)= &[A_{\rm raw}(D^+\to K_S^0\pi^+)(t_1,t_2)-A_{\rm raw}(D^+\to K^-\pi^+\pi^+)]\\\label{zaqwq}&-[A_{\rm raw}(D^+_s\to K_S^0K^+)(t_1,t_2)-A_{\rm raw}(D^+_s\to\pi^+K^+K^-)].
\end{align}
The dependences of $\Delta A_{CP}(\pi^+,K^+)$ on $t_1$ and $t_2$ are displayed in Fig.~\ref{fig:DeltaAcp}.
One can find that $\Delta A_{CP}(\pi^+,K^+)$ is of order of $10^{-3}$ in most time intervals.
Such magnitudes are available on Belle II and LHCb in the future and the favorable time intervals can be selected to facilitate measurements.

\section{Summary}\label{CON}

In this work, we systematically investigate the CP asymmetries in the $D\to K^0_{S,L}P$ and $D\to K^0_{S,L}V$ decays.
The formulas of time-dependent and time-integrated CP asymmetries are analyzed in detail.
The $D^0-\overline D^0$ mixing effects and the $K^0_L$ modes are considered for the first time.
The hadronic parameters are extracted within the topological diagram approach by fitting the branching fractions of the Cabibbo-favored and doubly Cabibbo-suppressed modes.
The tension between theoretical predictions and experimental data of $K^0_S-K^0_L$ asymmetries in the $D^0\to K_{S,L}^0\omega$ and $D^0\to K_{S,L}^0\phi$ modes is mitigated.
The interference between the Cabibbo-favored and doubly Cabibbo-suppressed amplitudes with the mixing of final-state mesons can reach to be $\mathcal{O}(10^{-3})$ order in the $D^+\to K^0_S\pi^+$, $D^+\to K^0_SK^+$, $D^0\to K^0_S\rho^0$, and $D^0\to K^0_S\phi$ modes.
The difference in CP asymmetries between the $D^+\to K^0_S\pi^+$ and $D^+_s\to K^0_SK^+$ modes is predicted to be available on LHCb and Belle II in the near future.

\begin{acknowledgements}
This work was supported in part by Scientific Research Fund of Hunan Provincial Department under No. 25B0090 and 24B0435 and the National Natural Science
Foundation of China under No. 32502553.
\end{acknowledgements}

\begin{appendix}
\addcontentsline{toc}{section}{Appendix}
\section{CP conjugate amplitudes}\label{cp}
With the convention \,$\mathcal{CP}|K^0\rangle = -|\overline{K}^0\rangle$ and the unitarity of the $\mathcal{P}$ and $\mathcal{C}$ operators, we get
\begin{equation}\label{zq8}
\begin{split}
\mathcal{A}(\overline{D}\to K^0\overline{f})  = \langle K^0\overline{f}|\mathcal{H}_w|\overline{D} \rangle
=\langle K^0\overline{f}|(\mathcal{CP})^\dagger(\mathcal{CP})\mathcal{H}_w(\mathcal{CP})^\dagger(\mathcal{CP})|\overline{D} \rangle
=-\langle \overline K^0f|\mathcal{H_{CP}}|D \rangle,
\end{split}
\end{equation}
where $\mathcal{H_{CP}}=(\mathcal{CP})\mathcal{H}_w(\mathcal{CP})^\dagger$ is the CP conjugate of $\mathcal{H}_w$.
$\mathcal{H}_w$ is the effective Hamiltonian of $|\Delta C|=1$ transitions, which can be written as
\begin{equation}
\mathcal{H}_w = \frac{G_F}{\sqrt{2}}\Big(V_{CKM}\sum_i^n\big(C_i(\mu) Q^{\Delta C = 1}_i(\mu)\big)+V^*_{CKM}\sum_i^n\big(C_i(\mu) Q^{\Delta C = -1}_i(\mu)\big)\Big),
\end{equation}
where $G_F$ denotes the Fermi coupling constant, $V_{CKM}$ is the product of the Cabibbo-Kobayashi-Maskawa (CKM) matrix elements, and $C_{1,2}(\mu)$ are the Wilson coefficients.
The CP transformation of the effective Hamiltonian is derived as
\begin{align}\nonumber
\mathcal{H_{CP}}& = (\mathcal{CP})\frac{G_f}{\sqrt{2}}\Big(V_{CKM}\sum_i^n\big(C_i(\mu) Q^{\Delta C = 1}_i(\mu)\big)+V^*_{CKM}\sum_i^n\big(C_i(\mu) Q^{\Delta C = -1}_i(\mu)\big)\Big)(\mathcal{CP})^\dagger
\\\nonumber
& =\frac{G_f}{\sqrt{2}}\Big(V_{CKM}\sum_i^n\big(C_i(\mu) ~(\mathcal{CP})Q^{\Delta C = 1}_i(\mu)(\mathcal{CP})^\dagger\big)+V^*_{CKM}\sum_i^n\big(C_i(\mu) ~(\mathcal{CP})Q^{\Delta C = -1}_i(\mu)(\mathcal{CP})^\dagger\big)\Big) \\\label{z3}
& =\frac{G_f}{\sqrt{2}}\Big(V_{CKM}\sum_i^n\big(C_i(\mu) Q^{\Delta C = -1}_i(\mu)\big)+V^*_{CKM}\sum_i^n\big(C_i(\mu) ~Q^{\Delta C = 1}_i(\mu)\big)\Big).
\end{align}
Here, we have used the relation
\begin{align}\label{qq}
(\mathcal{CP})Q_i^{\Delta C=\pm1}(\mu)(\mathcal{CP})^\dag=Q_i^{\Delta C=\mp1}(\mu).
\end{align}
Taking the effective operator of Cabibbo-favored decays $Q^{\Delta C = 1}_{CF}=\bar{u}\gamma_\mu(1-\gamma_5)d\otimes\bar{s}\gamma^\mu(1-\gamma_5)c$ as an example, we illustrate Eq.~\eqref{qq} explicitly.
The fermion bilinears $\bar\psi_1\gamma^\mu\psi_2$ and $\bar\psi_1\gamma^\mu\gamma_5\psi_2$ transform as follows \cite{Peskin:1995ev}:
\begin{equation}\label{zq9}
  (\mathcal{CP})(\bar\psi_1\gamma^\mu\psi_2)(\mathcal{CP})^\dag=-\bar\psi_2\gamma_\mu\psi_1, \qquad (\mathcal{CP})(\bar\psi_1\gamma^\mu\gamma_5\psi_2)(\mathcal{CP})^\dag=-\bar\psi_2\gamma_\mu\gamma_5\psi_1.
\end{equation}
Then we get
\begin{align}\nonumber
  (\mathcal{CP})\big(\bar{u}\gamma_\mu(1-\gamma_5)d\big)(\mathcal{CP})^\dag&=-\bar{d}\gamma^\mu(1-\gamma_5)u,\\\label{zq10} (\mathcal{CP})\big(\bar{s}\gamma^\mu(1-\gamma_5)c\big)(\mathcal{CP})^\dag&=-\bar{c}\gamma_\mu(1-\gamma_5)s.
\end{align}
The CP transformation of operator $Q^{\Delta C = 1}_{CF}(\mu)$ then can be written as
\begin{align}\nonumber
(\mathcal{CP}) Q^{\Delta C = 1}_{CF}(\mathcal{CP})^\dag&= (\mathcal{CP})\big(\bar{u}\gamma_\mu(1-\gamma_5)d\otimes\bar{s}\gamma^\mu(1-\gamma_5)c)(\mathcal{CP})^\dag
\\\label{zq11}
&=\bar{d}\gamma^\mu(1-\gamma_5)u\otimes\bar{c}\gamma_\mu(1-\gamma_5)s =Q^{\Delta C = -1}_{CF}.
\end{align}
By comparing $\mathcal{H}_w$ with $\mathcal{H_{CP}}$, one can find that the corresponding CKM elements of the operators $Q_i^{\Delta C =1}(\mu)$ and $Q_i^{\Delta C =-1}(\mu)$ are exchanged.
According to Eqs.~\eqref{eq:ampCFDCS}, ~\eqref{zq8} and \eqref{z3}, we get
\begin{equation}
  \mathcal{A}(\overline{D}\to K^0\overline{f}) = -\langle \overline{K}^0f|\mathcal{H_{CP}}|D \rangle = - \mathcal{T}_{CF}e^{i(-\phi_{CF}+\delta_{CF})}.
\end{equation}
With the similar derivation, we get
\begin{equation}
  \mathcal{A}(\overline{D}\to \overline{K}^0\overline{f}) = -\langle K^0f|\mathcal{H_{CP}}|D \rangle = - \mathcal{T}_{DCS}e^{i(-\phi_{DCS}+\delta_{DCS})}.
\end{equation}

\end{appendix}


%

\end{document}